\documentclass[11pt,letterpaper]{article}
\pdfoutput=1
\usepackage{jheppub}
\usepackage{color}
\usepackage{graphicx}

\usepackage{verbatim}
\usepackage{amsmath}
\usepackage{amssymb}
\usepackage{subfig}
\usepackage{url}
\usepackage{bbold}
\usepackage{slashed}

\usepackage{multirow}
\usepackage{threeparttable}
\usepackage{paralist}

\newcommand{\met}{\slashed{E}_{T}}
\newcommand{\mev}{\mathrm{MeV}}
\newcommand{\gev}{\mathrm{GeV}}
\newcommand{\tev}{\mathrm{TeV}}
\newcommand{\ifb}{\mathrm{fb}^{-1}}

\newcommand{\cm}{\mathrm{cm}}
\newcommand{\mm}{\mathrm{mm}}
\newcommand{\m}{\mathrm{m}}
\newcommand{\para}{\paragraph{}}

\begin{document}

\title{Neutralino Dark Matter at 14 TeV and 100 TeV}
\author{Matthew Low}
\author{and Lian-Tao Wang}
\affiliation{Department of Physics, Enrico Fermi Institute, and Kavli Institute for Cosmological Physics, University of Chicago, Chicago, IL 60637}
\emailAdd{mattlow@uchicago.edu}
\emailAdd{liantaow@uchicago.edu}

\date{\today}

\abstract{
In recent years the search for dark matter has intensified with competitive bounds coming from collider searches, direct detection, and indirect detection.  Collider searches at the Large Hadron Collider (LHC) lack the necessary center-of-mass energy to probe TeV-scale dark matter.  It is TeV-scale dark matter, however, that remains viable for many models of supersymmetry.  In this paper, we study the reach of a 100 TeV proton-proton collider for neutralino dark matter and compare to 14 TeV LHC projections.  We employ a supersymmetric simplified model approach and present reach estimates from monojet searches, soft lepton searches, and disappearing track searches.  The searches are applied to pure neutralino spectra, compressed neutralino spectra, and coannihilating spectra.  We find a factor of 4-5 improvement in mass reach in going from 14 TeV to 100 TeV.  More specifically, we find that given a 1\% systematic uncertainty, a 100 TeV collider could exclude winos up to 1.4 TeV and higgsinos up to 850 GeV in the monojet channel.  Coannihilation scenarios with gluinos can be excluded with neutralino masses of 6.2 TeV, with stops at 2.8 TeV, and with squarks at 4.0 TeV.  Using a soft lepton search, compressed spectra with a chargino-neutralino splitting of $\Delta m = 20 - 30$ GeV can exclude neutralinos at $\sim$ 1 TeV.  Given a sufficiently long chargino lifetime, the disappearing track search is very effective and we extrapolate current experimental bounds to estimate that a $\sim$ 2 TeV wino could be discovered and a $\sim$ 3 TeV wino could be excluded.}

%\keywords{dark matter, supersymmetry, neutralino}
%\arxivnumber{XXXX.XXXX}
\preprint{EFI {14-6}}
\maketitle

%%%%%%%%%%%%%%%%%%%%%%%%%%%%%%%%%%%%%%%%%%%%%%%%%%%%%%%%%%%%%%%%%%
%%%%%%%%%%%%%%%%%%%%%%%%%%%%%%%%%%%%%%%%%%%%%%%%%%%%%%%%%%%%%%%%%%
\section{Introduction}
\label{sec:intro}

The existence of cold dark matter is one of very few pieces of experiment evidence for physics beyond the standard model~\cite{Bertone:2004pz}.  Its identity, however, remains one of the most outstanding questions in particle physics.  Among the myriad possibilities, a weakly interacting massive particle (WIMP) is one of the most compelling.  The WIMP scenario assumes that dark matter has weak, but still sizable interactions with the standard model.  The cross-section for a pair of WIMPs interacting with a pair of standard model particles can be written as $\sigma \propto g_{\rm eff}^4 /M^2_{\rm DM} $, where $g_{\rm eff}$ is the effective coupling characterizing the interaction.  To avoid overclosing the universe, we must have

\begin{equation}
  M_{\rm DM}  \lesssim 1.8 \text{ TeV}  \left( \frac{g_{\rm eff}^2}{0.3}\right).
\label{eq:wimp_mass}
\end{equation}

Thus the WIMP mass is expected to be near the weak scale, which offers the exciting prospect of discovery at on-going or future collider experiments.

\para
It is certainly possible that dark matter is part of a complete TeV-scale new physics model which contains other new particles that do not comprise the dark matter content of the universe.  Low energy supersymmetry (SUSY) is one of the most prominent examples~\cite{Jungman:1995df,Martin:1997ns}.  In the usual naturalness-motivated SUSY scenarios, all of the superpartners are relatively light and the lightest superpartner (LSP) is stable and makes up the full dark matter content of the universe.  As stringent constraints already exist on the properties of dark matter, including it be neutral and uncolored, in the minimal supersymmetric standard model (MSSM) the identity of the LSP is the lightest neutralino.  At hadron colliders such as the Large Hadron Collider (LHC), the pair production of colored superpartners, squarks and gluinos, are the dominant SUSY processes.  The squarks or gluinos then decay both to standard model particles and the LSP via a, possibly long, decay chain.  The standard SUSY searches designed for this scenario look for jets and missing energy and would discover dark matter along with other superpartners~\cite{Alves:2011wf,Cohen:2013xda,Fowlie:2014awa}.  Thus the mass reach for dark matter is strongly correlated with that of gluinos and squarks.  Due to kinematics, the mass reach for the LSP is typically weaker by a few hundred GeV.

\para
Concurrently with complete models, it is also crucial to explore the discovery potential for dark matter independent of the presence of additional new particles.  The starting point here would be to consider a set of simple examples of possible dark matter candidates.  Since dark matter cannot be electrically charged, the simplest possibility is a dark matter particle as the neutral member of a standard model weak multiplet.  WIMPs, as their name suggests, fall into this category.  Neutralino dark matter, in fact, already provides concrete realizations of three of the simplest cases.  The higgsino is a vector-like doublet, the wino is a triplet, and the bino provides an example of a singlet which can mix with the other multiplets after electroweak symmetry breaking.  Therefore, as a first step to directly studying dark matter, we choose to consider dark matter as a neutralino LSP.  In general, except in coannihilation scenarios, we will assume all other squarks and sleptons are too heavy to be relevant for dark matter production at colliders.  These have been called simplified models of dark matter and are similar to the split SUSY scenario~\cite{Wells:2003tf,ArkaniHamed:2004fb,Giudice:2004tc}.

As we employ simplified models, the basic collider process is simply dark matter pair production.  As dark matter will escape the detector unseen, there needs to be additional hard radiation of a standard model particle, which could be a quark or gluon~\cite{Beltran:2010ww,Fox:2011pm}, photon~\cite{Gershtein:2008bf,Fox:2011fx}, $W$~\cite{Bai:2012xg} or $Z$~\cite{Petriello:2008pu,Carpenter:2012rg}, or even a higgs~\cite{Petrov:2013nia,Carpenter:2013xra,Berlin:2014cfa}.  Among them, a quark or gluon emission, the {\it monojet} channel, is typically the most sensitive.  When the LSP is a mixed state, the LSP can be separated by a mass splitting $\Delta m \equiv m_{\tilde{\chi}^\pm_1} - m_{\tilde{\chi}^0_1}\sim 20-50~\gev$ from the other chargino and neutralino states.  In this case, in addition to a hard jet, it is possible to search for low $p_T$ leptons resulting from a chargino or neutralino which decays to the LSP and leptons or quarks.  We call this the {\it soft lepton} channel.  When the LSP is a pure state, at tree-level it is mass degenerate with the charged components.  As the mass splitting is generated at loop-level, it is small and can lead to charginos with a macroscopic lifetime.  This leaves a rather striking signature of high $p_T$ charged track abruptly ending when the chargino decays to the LSP and very soft, likely undetected, standard model particles.  We also include this {\it disappearing tracks} search in our consideration.

\para
Since typically only a small fraction of the center of mass energy of a hadron collider is available in hard collisions, we expect only a small part of the WIMP parameter space, shown in Eq.~\ref{eq:wimp_mass}, can be probed at the $14~\tev$ LHC.  This has been confirmed by a number of recent studies~\cite{Schwaller:2013baa,Han:2013usa,Bhattacherjee:2013wna,Baer:2014cua,Han:2014kaa}.  In our study, we focus on the prospects at a $100~\tev$ proton-proton collider~\cite{Zhou:2013raa,Cohen:2013xda}.  For comparison, we will also often show results for the $14~\tev$ LHC.

We briefly outline our simulation procedure and sketch the analyses implemented in Sect.~\ref{sec:overview}.  We then present results for pure wino dark matter and pure higgsino dark matter in Sect.~\ref{sec:wino} and Sect.~\ref{sec:higgsino}, respectively.  Mixed scenarios are shown in Sect.~\ref{sec:mixed}, as well as coannihilation spectra in Sect.~\ref{sec:coan}.  Finally our conclusions are summarized in Sect.~\ref{sec:conclusions}.  The first appendix, App.~\ref{app:analysis}, provides a more comprehensive description of the analyses and the second appendix, App.~\ref{app:systematics}, includes a brief discussion on the dominant systematic uncertainties.

%%%%%%%%%%%%%%%%%%%%%%%%%%%%%%%%%%%%%%%%%%%%%%%%%%%%%%%%%%%%%%%%%%
%%%%%%%%%%%%%%%%%%%%%%%%%%%%%%%%%%%%%%%%%%%%%%%%%%%%%%%%%%%%%%%%%%
\section{Analysis Overview}
\label{sec:overview}

In this section we give a brief overview of how we simulated events and the various analyses applied.  A more thorough discussion is found in App.~\ref{app:analysis}.  The SUSY mass spectra were computed with Suspect2~\cite{Djouadi:2002ze} using parameters input at the weak-scale with $\tan\beta=20$.  Events were generated using MadGraph 5 v1.5.12~\cite{Alwall:2011uj} for matrix elements and Pythia 6~\cite{Sjostrand:2006za} for showering and hadronization.  All signals and backgrounds were matched up to two additional jets using the MLM matching scheme (except for $t\bar{t}$ which was matched to one additional jet).  Delphes 3 v3.0.9~\cite{deFavereau:2013fsa} was used as a detector simulation with the Snowmass detector card~\cite{Anderson:2013kxz,Avetisyan:2013dta,Avetisyan:2013onh}.  The Snowmass Delphes settings identify electrons and muons above $10~\gev$, tag hadronic tau leptons with an efficiency of $65\%$, and cluster anti-$k_T$ jets~\cite{Cacciari:2008gp} with a radius of $R=0.5$ using Fastjet v3.0.3~\cite{Cacciari:2011ma}.

The inclusive pair production cross-sections were verified using Prospino 2~\cite{Beenakker:1999xh} for pure spectra.  While the k-factors are known for the $2 \rightarrow 2$ process, a collider search requires there be a hard visible object to trigger on corresponding to the $2 \to 3$ process.  The k-factor for this process is not the same as the $2 \rightarrow 2$ process.  In fact, since going to next-to-leading order from leading-order opens up different new partonic channels for the $2 \rightarrow 2$ and $2 \rightarrow 3$ processes, we should expect the pair production k-factor to differ from k-factor for the monojet search~\cite{Cullen:2012eh}.  In this study we choose not to apply k-factors on the signal or background.  As the signal cross-section falls very quickly with mass, we expect neglecting k-factors to only introduce a small error.

\subsection*{Monojet}

The first analysis we implement looks for one hard jet produced with large missing energy ($\met$), known as a monojet search.  Monojet searches have been carried out both at the Tevatron~\cite{Abazov:2003gp,Aaltonen:2008hh,Abazov:2008kp} and the LHC~\cite{ATLAS:2012ky,Chatrchyan:2012me,ATLAS:2012zim,CMS:2013rwa} looking both for large extra dimensions and dark matter via a contact operator~\cite{Beltran:2010ww,Fox:2011pm} or light mediator~\cite{An:2012ue,An:2012va,An:2013xka}.  We follow the format of the most recent CMS analysis which uses $19.5~\ifb$ of $8~\tev$ data~\cite{CMS:2013rwa}.  The CMS search requires one or two hard jets, where the second jet cannot be back-to-back with the first jet, and large $\met$.  Events that contain electrons, muons, or tagged hadronic taus are vetoed.

The backgrounds for this channel include standard model processes with a hard jet and neutrinos.  Despite the lepton veto, processes with leptons also comprise part of the background because leptons can fail to be tagged if they are outside the detector acceptance, not isolated, or too soft.  This leads to a long list of backgrounds: $Z(\nu\nu)+\text{jets}$, $W(\ell\nu)+\text{jets}$, $t\bar{t}$, $Z(\ell\ell)+\text{jets}$, single $t$, diboson production, and QCD multijets\footnote{While QCD multijet events do not contain real missing energy, mismeasured jets can fake missing energy.  As the multijet rate is dominated by dijet events requiring events with two jets to not be back-to-back effectively removes the QCD background.}.  Of these, only $Z(\nu\nu)+\text{jets}$ is irreducible and it comprises roughly $70\%$ of the background in the signal region.  Together with $W(\ell\nu)+\text{jets}$ these backgrounds make up $99\%$ of the background.  For completeness we generate 
$Z(\nu\nu)+\text{jets}$, $W(\ell\nu)+\text{jets}$, $t\bar{t}$, $Z(\ell\ell)+\text{jets}$, and $W(\ell\nu)W(\ell\nu)+\text{jets}$ backgrounds with MadGraph.

\subsection*{Soft Leptons}

When the LSP is split from other electroweakino states by $\Delta m = 20-50~\gev$ then these states can also be pair produced and decay to the LSP via off-shell gauge bosons which decay hadronically or into low $p_T$ leptons.  The hadronic decays are difficult to extricate from the busy hadronic environment, but it is possible to tag the soft leptons.  This is different from the standard multilepton searches~\cite{CMS:2013dea,CMS:2013jfa,CMS:2013ida,Aad:2014nua,Aad:2014vma} where there are both more and harder leptons.  It has been noted in~\cite{Giudice:2010wb,Gori:2013ala} that triggering on a hard jet, as in the monojet search, is advantageous in a soft lepton search.  As the optimal search strategy strongly depends on the electroweakino splittings, it would be interesting to look at the transition between a pure monojet search yielding the highest significance and a traditional multilepton search yielding the highest significance.  This is beyond the scope of this note, and we restrict ourselves to the compressed region where $\Delta m = 20-30~\gev$.

As in the monojet search, events are required to have one or two jets that are not back-to-back and large missing energy.  Rather than applying a lepton veto, events are binned according to whether they contain 0, 1, or 2 soft leptons.  The significances are computed separately in each lepton bin and added in quadrature.  The backgrounds are the same as the monojet search, but the background composition varies significantly across lepton bins.  Like the monojet channel, the 0-lepton bin is about $99\%$ composed of $Z(\nu\nu)+\text{jets}$ and $W(\ell\nu)+\text{jets}$.  The 1-lepton bin is more than $90\%$ from $W(\ell\nu)+\text{jets}$ and the 2-lepton bin is dominated by $W(\ell\nu)W(\ell\nu)+\text{jets}$ at roughly $70 \%$ with the rest coming from $Z(\ell\ell)+\text{jets}$ and $t\bar{t}$.

\subsection*{Disappearing Tracks}

The third analysis leverages the fact that in scenarios with dark matter as a pure state, a chargino are often near-degenerate with the LSP.  In the limit $m_Z/m_{\tilde{\chi}} \to 0$, a pure wino has a splitting of $\approx 166~\mev$ and a pure higgsino has a splitting of $\approx 355~\mev$~\cite{Thomas:1998wy,Cirelli:2005uq}.  Due to the small mass splitting, the dominant decay $\tilde{\chi}^\pm \to \pi^\pm + \tilde{\chi}^0$ has a long lifetime.  Thus, a fraction of the charginos can live long enough, $c\tau \sim 6~\cm$, to leave a track in the inner detector.  A number of phenomenological studies have been done~\cite{Feng:1994mq,Feng:1996ag,Thomas:1998wy,Feng:1999fu,Gunion:1999jr,Gunion:2001fu,Barr:2002ex,Ibe:2006de,Buckley:2009kv}.  This is a promising search channel with no obvious physics background.  One possibility is to look for so-called disappearing tracks, in which a chargino decays in the inner detector, resulting in a track that disappears where the chargino decays into a neutralino and a soft pion\footnote{The signature has also been called {\it kinked tracks} or {\it track stubs}.  It is worth noting that this signal is part of a larger class of signatures of particles that traverse macroscopic distances before decaying.  While it is detector-dependent, roughly speaking charged particles with a lifetime $c\tau = \mathcal{O}(\mm)$ result in displaced vertices, charged particles with a lifetime $c\tau = \mathcal{O}(\cm)$ result in disappearing tracks, and charged particles with a lifetime $c\tau = \mathcal{O}(\m)$ result in stable charged massive particles.}.

We derive our projections from a recent ATLAS search that reported a $95 \%$ exclusion limit close to $250~\gev$, using $20.3~\ifb$ of $8~\tev$ data~\cite{Aad:2013yna}.  Similar to the monojet analysis, this search triggers on a hard jet and large $\met$, additionally requiring a disappearing track.  While the monojet analysis has not yet reached the sensitivity necessary to probe the pure wino or pure higgsino scenarios, the disappearing track search is already starting to exclude regions of the pure wino parameter space.  Therefore it is reasonable to expect that this channel will be much stronger both in the $14~\tev$ LHC run and at a $100~\tev$ proton-proton collider.

\para
The significance of a given search is calculated as
\begin{equation}
  \text{Significance} = \frac{S}{\delta B} = \frac{S}{\sqrt{B + \lambda^2 B^2 + \gamma^2 S^2}} ,
  \label{eq:signif}
\end{equation}
where $\lambda$ and $\gamma$ parameterize the systematic uncertainty on the background and on the signal, respectively.  While we assume the systematics are the same across background channels, considering different systematics for each background would not noticeably change the results, as each search is dominated by one or two backgrounds.

Our analyses have not included effects from pileup.  As a future high energy proton-proton collider will likely operate with high instantaneous luminosity, events will contain a high level of hadronic contamination from pileup.  In a fully realistic projection it is important to consider the effects of pileup and the effects of applying the appropriate pileup removal techniques~\cite{Cacciari:2007fd,Krohn:2013lba,Berta:2014eza}.  For the analyses presented in this paper, however, events are selected with a very hard cut on the leading jet and missing energy so we expect such additional considerations will not significantly alter the results.

%%%%%%%%%%%%%%%%%%%%%%%%%%%%%%%%%%%%%%%%%%%%%%%%%%%%%%%%%%%%%%%%%%
%%%%%%%%%%%%%%%%%%%%%%%%%%%%%%%%%%%%%%%%%%%%%%%%%%%%%%%%%%%%%%%%%%
\section{Pure Wino}
\label{sec:wino}

The first set of SUSY spectra we consider are those with a pure wino LSP.  This scenario can be realized if anomaly mediation the main mechanism through which the gaugino soft masses are generated~\cite{Randall:1998uk,Giudice:1998xp}.  Models which implement this, along with the feature that the scalar are heavy (compared to the gravitino mass) include split SUSY~\cite{Wells:2003tf,ArkaniHamed:2004fb,Giudice:2004tc,Cheung:2005ba}, mini-split susy~\cite{Arvanitaki:2012ps}, and spread susy~\cite{Hall:2011jd,Hall:2012zp}.

For a wino LSP to thermally saturate the relic density, it must have a mass of $m_{\tilde{\chi}} \sim 3.1~\tev$ (including the Sommerfeld effect)~\cite{Hisano:2006nn,Cohen:2013ama,Fan:2013faa}.  Assuming an NFW halo profile, current indirect detection experiments like Fermi~\cite{Ackermann:2011wa} and HESS~\cite{Abramowski:2013ax} constrain thermally produced winos to have a mass $m_{\tilde{\chi}} \lesssim 1.6~\tev$\footnote{Thermally produced winos with a mass $m_{\tilde{\chi}} \lesssim 3.1~\tev$ would only comprise part of the relic abundance.}.  Future independent detection experiments, like CTA, could move this bound down to $m_{\tilde{\chi}} \lesssim 1.1~\tev$~\cite{Consortium:2010bc,Bergstrom:2012vd}.  These limits, however, are subject to a number of astrophysics uncertainties.  Choosing different halo profile can move the HESS limit as low as $m_{\tilde{\chi}} \sim 0.5~\tev$ and as high as $m_{\tilde{\chi}} \sim 2.2~\tev$~\cite{Cohen:2013ama}.  Non-thermally produced, but relic density saturating, winos are ruled out across the parameter space up to $m_{\tilde{\chi}} \lesssim 25~\tev$.

Direct detection is another avenue through winos could be discovered.  In the heavy wino limit, the spin-independent scattering cross-section has been calculated to be $\sigma_{\text{SI}} = 1.3 \times 10^{-47}~\cm^2$~\cite{Hill:2013hoa}.  Future experiments are projected to probe this cross-section for dark matter masses of a few hundred GeV~\cite{Cushman:2013zza}.  TeV-scale dark matter is not only beyond the predicted reach, but also sits along the neutrino coherent scattering floor~\cite{Cushman:2013zza}.

As direct detection cannot probe thermally-saturating winos and indirect detection involves astrophysics uncertainties, there is a potentially interesting window in parameter space left open.  As will be shown, the LHC will not be able to cover it, as it is only sensitive to $m_{\tilde{\chi}} \sim 280-380~\gev$ winos.  A $100~\tev$ collider, on the other hand, may be able to reach $1.4-2.9~\tev$ and cover the parameter space.

\para
The wino is an electroweak triplet which results in one neutral and one charged state at low energies.  The pair production of charginos proceeds via the Drell-Yan-like process of an $s$-channel $Z$ going to a pair of charginos, which subsequently decay to the LSP and soft standard model particles.  Charginos can also be produced directly along with a neutralino via an $s$-channel $W^\pm$.

%%%%%%%%%%%%%%%%%%
\begin{figure}[h!]
  \centering
  \includegraphics[scale=0.5]{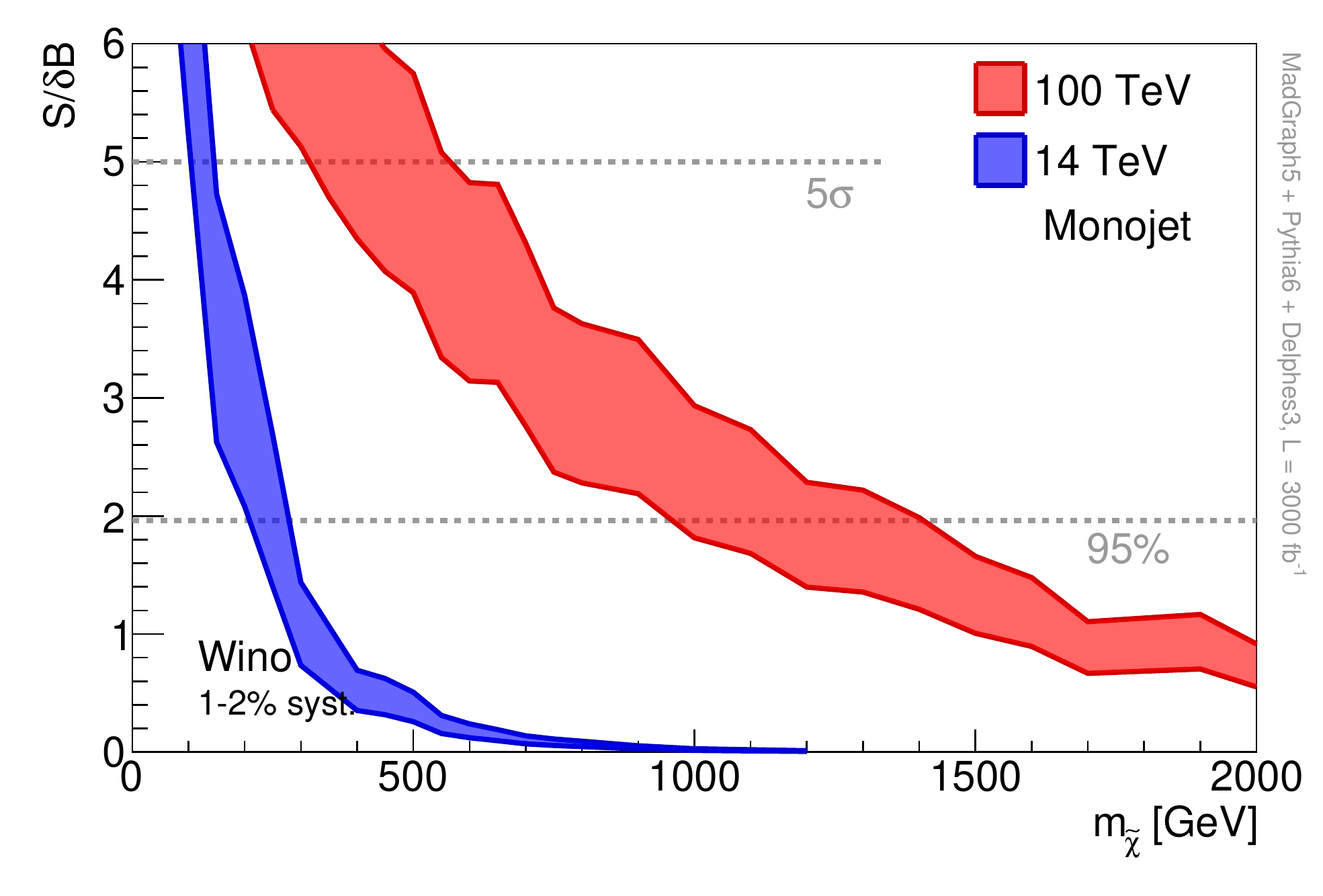}
  \caption{The mass reach in the pure wino scenario in the monojet channel with $\mathcal{L}=3000~\ifb$ for the $14~\tev$ LHC (blue) and a $100~\tev$ proton-proton collider (red).  The bands are generated by varying the background systematics between $1-2 \%$ and the signal systematic uncertainty is set to $10 \%$.}
  \label{fig:monojet_wino}
\end{figure}
%%%%%%%%%%%%

%The mass reach in the monojet channel for a pure wino LSP is shown in Fig.~\ref{fig:monojet_wino}.  The significance is computed using Eq.~\ref{eq:signif}.  We set $\gamma = 10 \%$, although its value has a very small effect since $S \ll B$ in much of the parameter space we explore.  The value of $\lambda$ is varied between $1-2 \%$, which generates the bands in the plot.  While the impact of background systematic uncertainty is clear, it is difficult to exactly project the systematic uncertainty of a future detector.  Naively scaling the current systematics in~\cite{CMS:2013rwa} with luminosity would result in $\approx 0.5 \%$ for $3000~\ifb$.  This is likely much too optimistic as components of the systematic uncertainty do not scale with statistics.  While even $1-2 \%$ may be overly optimistic, this at least provides a benchmark.  Given the importance of the monojet channel in the search for dark matter, not just in the pure wino case, minimizing possible systematics is a salient factor to be included in the design of detectors at future hadron colliders.
The mass reach in the monojet channel for a pure wino LSP is shown in Fig.~\ref{fig:monojet_wino}.  The significance is computed using Eq.~\ref{eq:signif}.  We set $\gamma = 10 \%$, although its value has a very small effect since $S \ll B$ in much of the parameter space we explore.  The value of $\lambda$ is varied between $1-2 \%$, which generates the bands in the plot.  In the large background limit Eq.~\ref{eq:signif} is $\approx (1/\lambda)(S/B)$, which means varying $\lambda$ by a factor of 2 can vary the significance by up to a factor of 2.  While the impact of background systematic uncertainty is clear, it is difficult to exactly project the systematic uncertainty of a future detector.  Naively scaling the current systematics in~\cite{CMS:2013rwa} with luminosity would result in $\approx 0.5 \%$ for $3000~\ifb$.  This is likely much too optimistic as components of the systematic uncertainty do not scale with statistics.  While even $1-2 \%$ may be overly optimistic, this at least provides a benchmark.  Given the importance of the monojet channel in the search for dark matter, not just in the pure wino case, minimizing possible systematics is a salient factor to be included in the design of detectors at future hadron colliders.

For reference, ignoring all systematics, at 14 TeV winos could be excluded at $m_{\tilde{\chi}} \sim 530~\gev$ and discovered at $m_{\tilde{\chi}} \sim 380~\gev$.  At 100 TeV the exclusion reach would be $m_{\tilde{\chi}} \sim 1.8~\tev$ and the discovery reach would be $m_{\tilde{\chi}} \sim 1.0~\tev$.

%%%%%%%%%%%%%%%%%%
\begin{figure}[h!]
  \centering
  \includegraphics[scale=0.35]{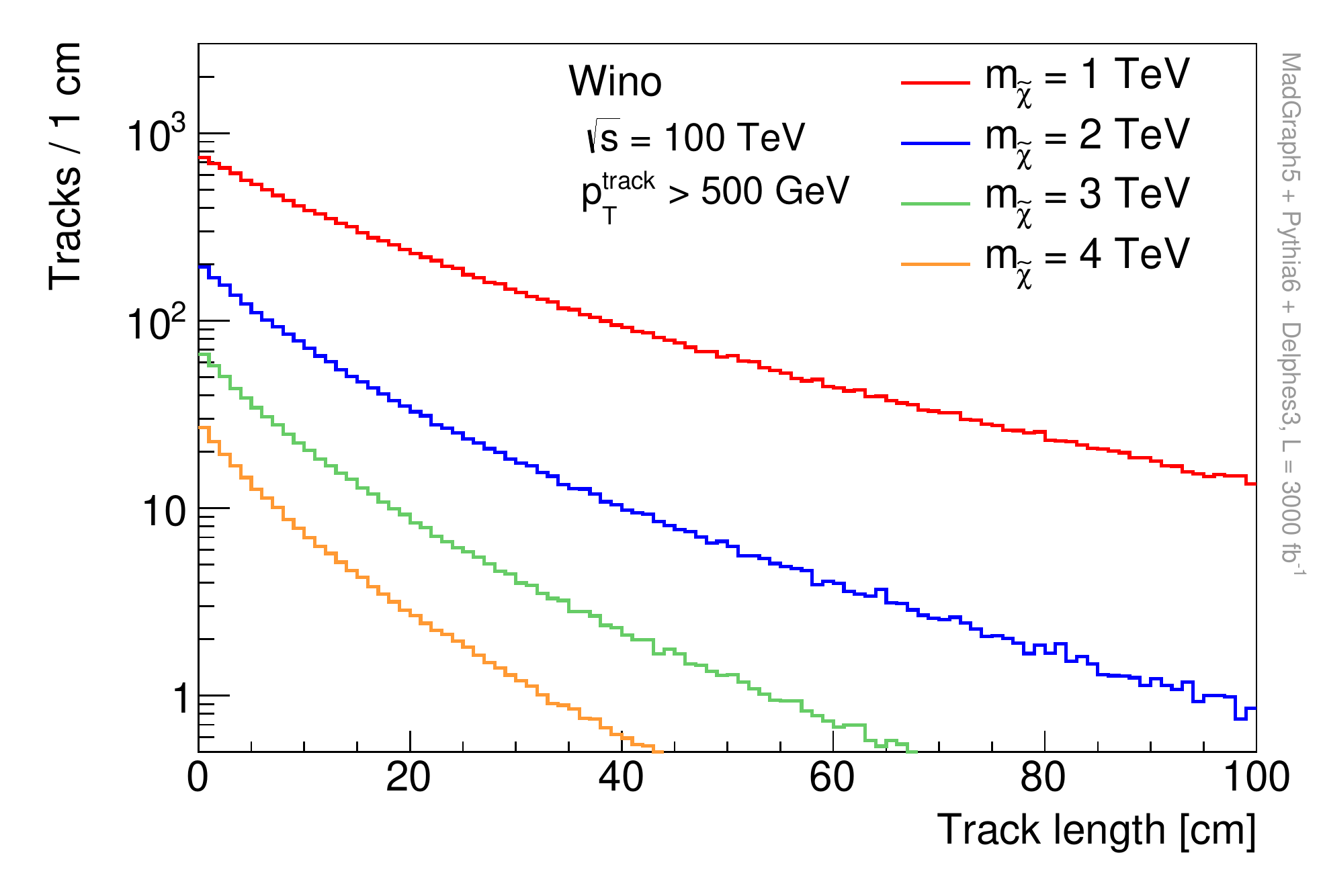} \includegraphics[scale=0.35]{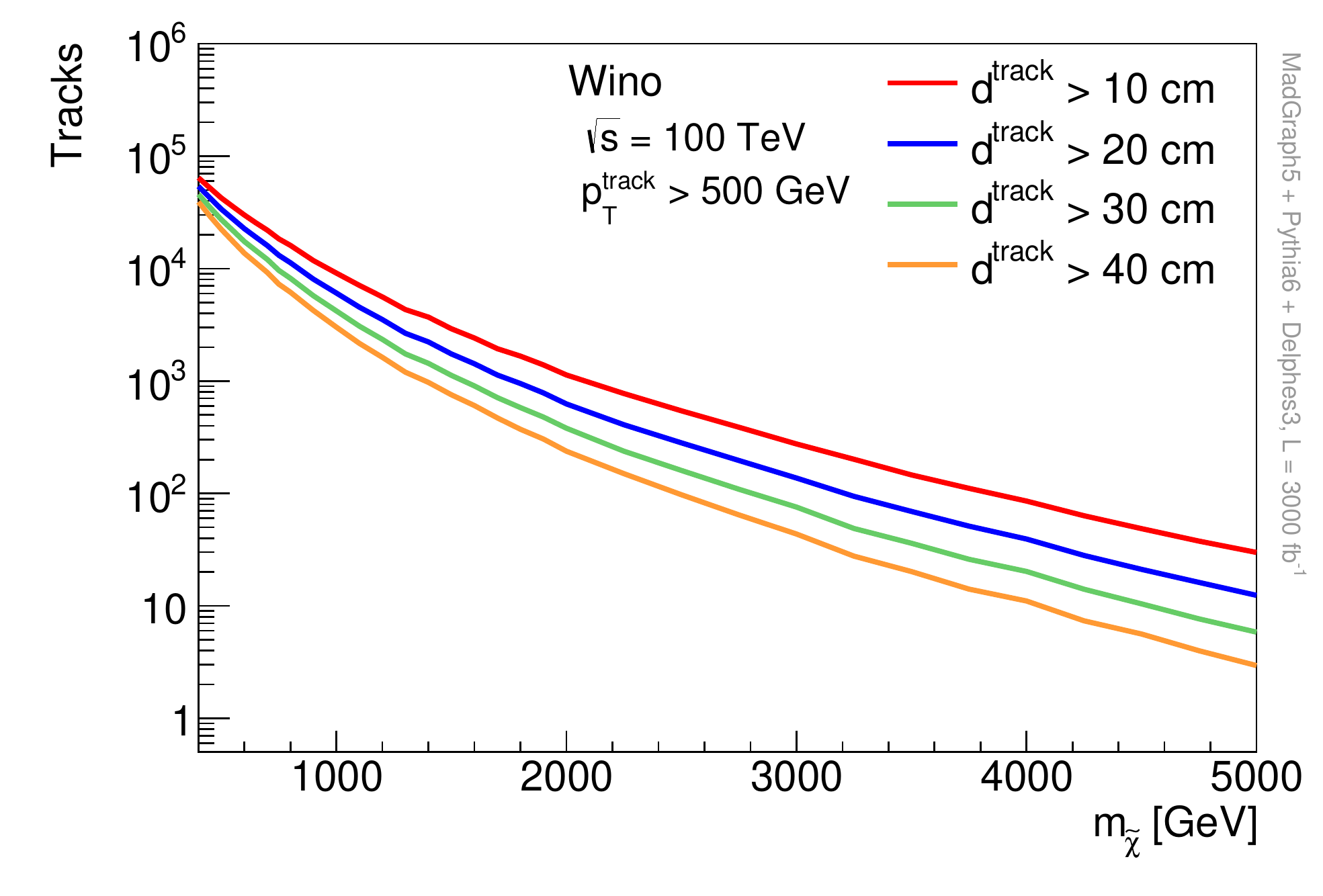}
  \caption{Chargino track distributions for the pure wino scenario showing the number of tracks for a given track length (left) and the number of tracks for a given wino mass (right).  Only events passing the analysis cuts in App.~\ref{app:analysis} and containing at least one chargino track with $p_T>500~\gev$ are considered.}
  \label{fig:track_distributions_wino}
\end{figure}
%%%%%%%%%%%%

\para
As mentioned, in the pure wino scenario, the mass splitting between the chargino and neutralino is generated by loop effects.  The value of the splitting has been calculated at two-loops to be $\Delta = 164.6~\mev$ in the large mass limit~\cite{Ibe:2012sx}, though the mass splitting varies very little with respect to wino mass.  A mass splitting can also be generated by higher dimension operators.  For the pure wino, the lowest operator that can split the charged and neutral states is dimension 7, so the splitting cited above is fairly model-independent.  In our simulation we use the lifetime calculated at one-loop.  At a collider the lifetime in the lab-frame also includes the velocity $\beta$ and boost $\gamma$ so that $d = \beta\gamma c\tau$.  Notice that $\beta\gamma$ can be substantially larger at $100~\tev$ than at $14~\tev$.  

The distribution of chargino track lengths is shown in Fig.~\ref{fig:track_distributions_wino} (left).  At ATLAS the disappearing track search is conducted using the tracker which has a high efficiency for selecting disappearing tracks starting at $d^{\text{track}} \sim 30~\cm$.  A detector with a similarly designed tracker would observe a handful of tracks for WIMPs as heavy as $m_{\tilde{\chi}} \sim 3~\tev$.  Fig.~\ref{fig:track_distributions_wino} (right) shows directly the number of tracks for a given LSP mass for various requirements on the length of a track.  While no upper limit on track length is enforced in Fig.~\ref{fig:track_distributions_wino}, as the distribution is exponential the value of the upper limit, $d^{\text{track}} \sim 80~\cm$ for ATLAS~\cite{Aad:2013yna}, has a negligible impact\footnote{The pure wino scenario results in a chargino lifetime of $c\tau \sim 6~\cm$ in the bulk of the mass range.  Even with the boost $d^{\text{track}} = \gamma\beta c\tau$, most charginos decay before reaching the end of the inner detector.  However, if the chargino lifetime were modified such that $c\tau \sim d^{\text{tracker}}$, then the length of the tracker becomes a relevant parameter.}.

%%%%%%%%%%%%%%%%%%
\begin{figure}[h!]
  \centering
  \includegraphics[scale=0.5]{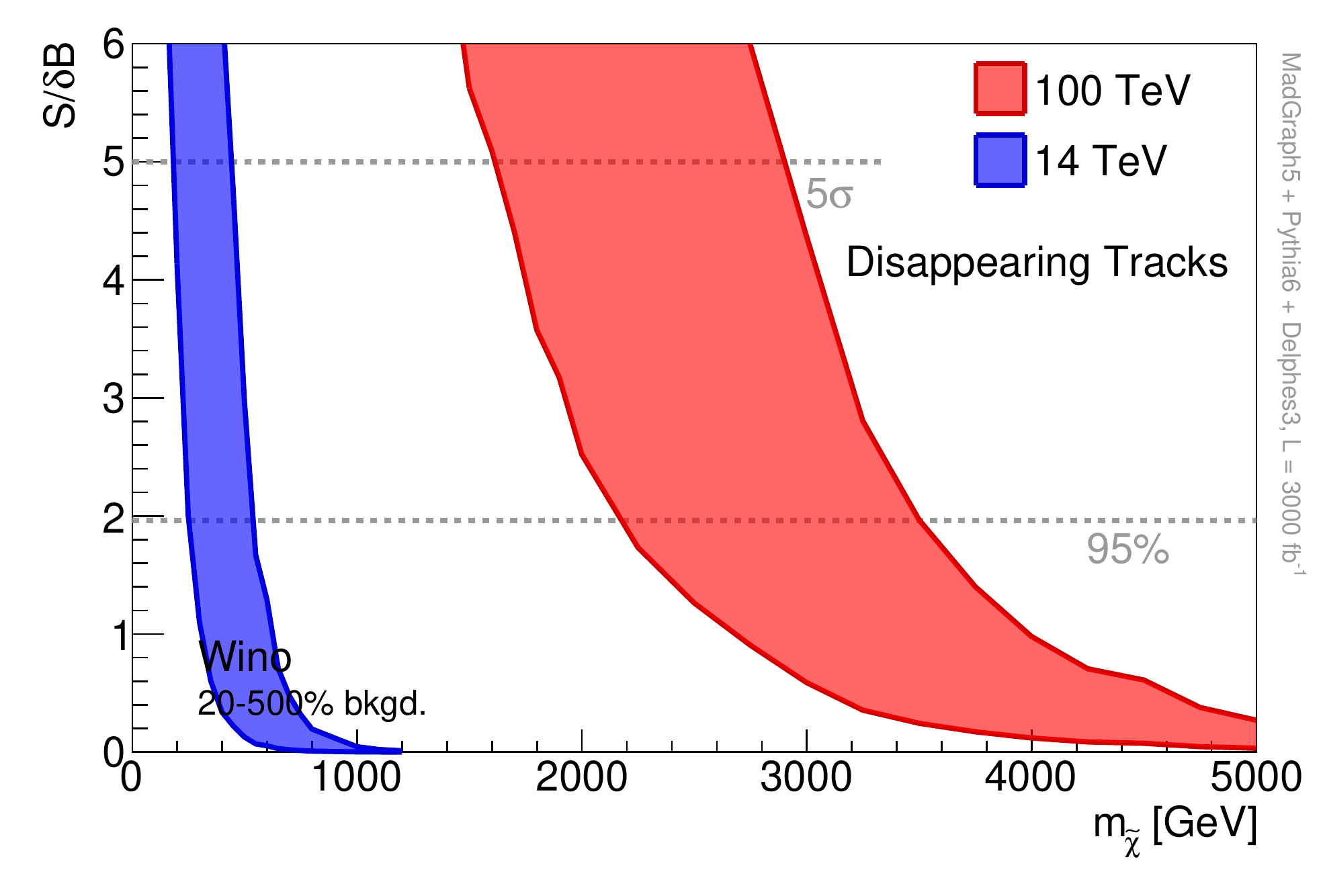}
  \caption{The mass reach in the pure wino scenario in the disappearing track channel with $\mathcal{L}=3000~\ifb$ for the $14~\tev$ LHC (blue) and a $100~\tev$ proton-proton collider (red).  The bands are generated by varying the background normalization between $20-500 \%$.  Only events passing the analysis cuts in App.~\ref{app:analysis} are considered.}
  \label{fig:track_wino}
\end{figure}
%%%%%%%%%%%%

Since the dominant background for a disappearing track search would be mismeasured low $p_T$ tracks, it is not possible to accurately project the background in a yet-to-be-designed detector at a 100 TeV proton-proton collider.  Nevertheless, Fig.~\ref{fig:track_distributions_wino} can serve as a rough guide.  For example, one could require $d^{\text{track}} > 30~\cm$ and there be tens of signal events passing all cuts, which is roughly where the $8~\tev$ ATLAS limit is set.  We choose to attempt a more systematic approach and naively extrapolate the dominant ATLAS background of mismeasured tracks.  The ATLAS search selects events with one or two hard jets and large $\met$ where neither of the jets can be too close to the $\met$ direction.  As this is the same criteria as the monojet search we estimate the background normalization to be set by the $Z(\nu\nu)+\text{jets}$ rate.  Additional details on our scaling procedure are found in App.~\ref{app:analysis}.  The results of the extrapolation are shown in Fig.~\ref{fig:track_wino} with $\gamma = 10 \%$ and $\lambda = 20 \%$.  The band is generated by varying the background normalization up and down by a factor of 5.

The results are summarized in Table~\ref{table:wino}.  In the monojet channel, we find that a $100~\tev$ collider extends the wino mass reach about $4-5$ times that of the LHC entering the $\tev$ mass range.  A much more promising search, however, is the disappearing track search.  Already at $8~\tev$ this channel has been shown to be more sensitive than a monojet search~\cite{Aad:2013yna}, and this continues to be the case at $100~\tev$.  Depending on the detector-backgrounds, this search has the potential to rule out (or perhaps discover) thermal winos.

%%%%%%%%%%%%%%%%%%
\begin{table}[h!]
  \centering
  \begin{tabular}{|c|c|c|c|c|c|}
    \hline
    \multirow{2}{*}{channel} & systematics/ & \multicolumn{2}{c|}{14 TeV} & \multicolumn{2}{c|}{100 TeV} \\ \cline{3-6}
                                         & normalization & $95 \%$ limit  & $5\sigma$ discovery & $95 \%$ limit  & $5\sigma$ discovery   \\ 
    \hline \hline
    \multirow{2}{*}{monojet}             & $1\%$    & $280~\gev$     & $140~\gev$   & $1.4~\tev$  & $560~\gev$            \\
                                         & $2\%$    & $205~\gev$     & $100~\gev$   & $960~\gev$  & $310~\gev$            \\ \hline
    \multirow{3}{*}{disappearing tracks} & $500\%$  & $250~\gev$     & $180~\gev$   & $2.1~\tev$  & $1.6~\tev$            \\
                                         & $100\%$  & $385~\gev$     & $295~\gev$   & $2.9~\tev$  & $2.2~\tev$            \\ 
                                         & $20\%$   & $535~\gev$     & $440~\gev$   & $3.5~\tev$  & $2.9~\tev$            \\ \hline
  \end{tabular}
  \caption{Mass reach for the pure wino scenario.  For the monojet channel, the second column shows the systematic uncertainty on the background used, while the systematic uncertainty on the signal was $10 \%$.  For the disappearing tracks channel, the second column shows the background normalization.  For this channel the background systematic uncertainty was $20 \%$ and the signal systematic uncertainty was $10 \%$.}
  \label{table:wino}
\end{table}
%%%%%%%%%%%%%%%%%%

%%%%%%%%%%%%%%%%%%%%%%%%%%%%%%%%%%%%%%%%%%%%%%%%%%%%%%%%%%%%%%%%%%
%%%%%%%%%%%%%%%%%%%%%%%%%%%%%%%%%%%%%%%%%%%%%%%%%%%%%%%%%%%%%%%%%%
\section{Pure Higgsino}
\label{sec:higgsino}

Another interesting class of SUSY spectra are those that contain a higgsino as the LSP.  Because of the connection between $\mu$ and fine-tuning, these spectra arise in natural SUSY~\cite{Hall:2011aa,Papucci:2011wy}, as well as in split SUSY~\cite{ArkaniHamed:2005yv} and mini-split SUSY~\cite{Arvanitaki:2012ps}.  A thermal higgsino saturates the relic density for $m_{\tilde{\chi}} \sim 1~\tev$, which like the thermal wino, is inaccessible to the LHC.  The spin-independent scattering cross-section has been calculated to be $\sigma_{\text{SI}} \lesssim 10^{-48}~\cm^2$ which is near or below the neutrino coherent scattering floor~\cite{Hill:2013hoa,Cushman:2013zza}.  While a $100~\tev$ collider can come much closer to the thermal value, likely it is still not able to rule out this scenario.

\para
The higgsino is a vector-like doublet which results in two neutralinos and one chargino at low energies.  This opens up additional pair production channels relative to the pure wino case, but all channels are still through an $s$-channel $W^\pm$ or $Z$.

%%%%%%%%%%%%%%%%%%
\begin{figure}[h!]
  \centering
  \includegraphics[scale=0.5]{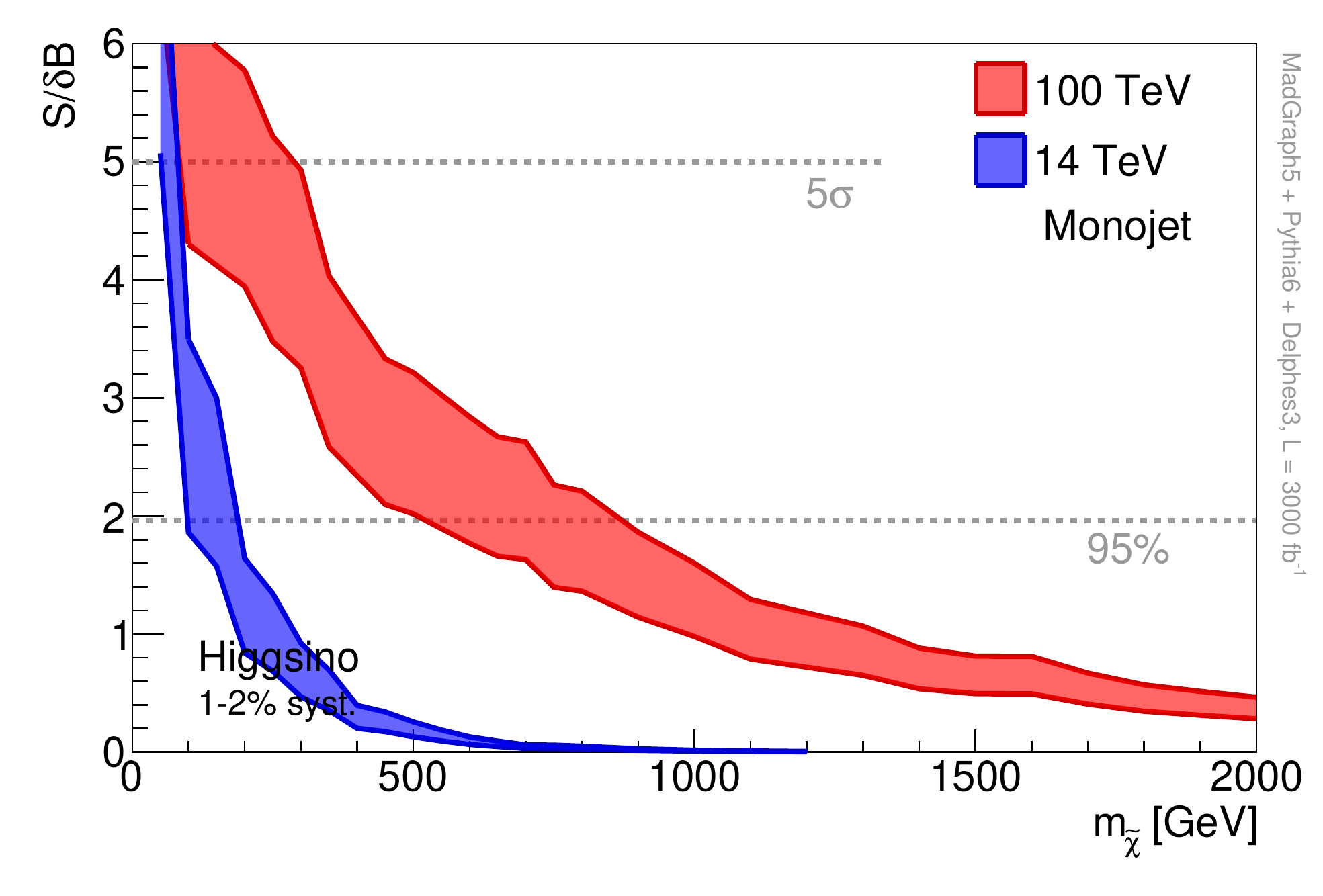}
  \caption{The mass reach in the pure higgsino scenario in the monojet channel with $\mathcal{L}=3000~\ifb$ for the $14~\tev$ LHC (blue) and a $100~\tev$ proton-proton collider (red).  The bands are generated by varying the background systematics between $1-2 \%$ and the signal systematic uncertainty is set to $10 \%$.}
  \label{fig:monojet_higgsino}
\end{figure}
%%%%%%%%%%%%

%%%%%%%%%%%%%%%%%%
\begin{figure}[h!]
  \centering
  \includegraphics[scale=0.35]{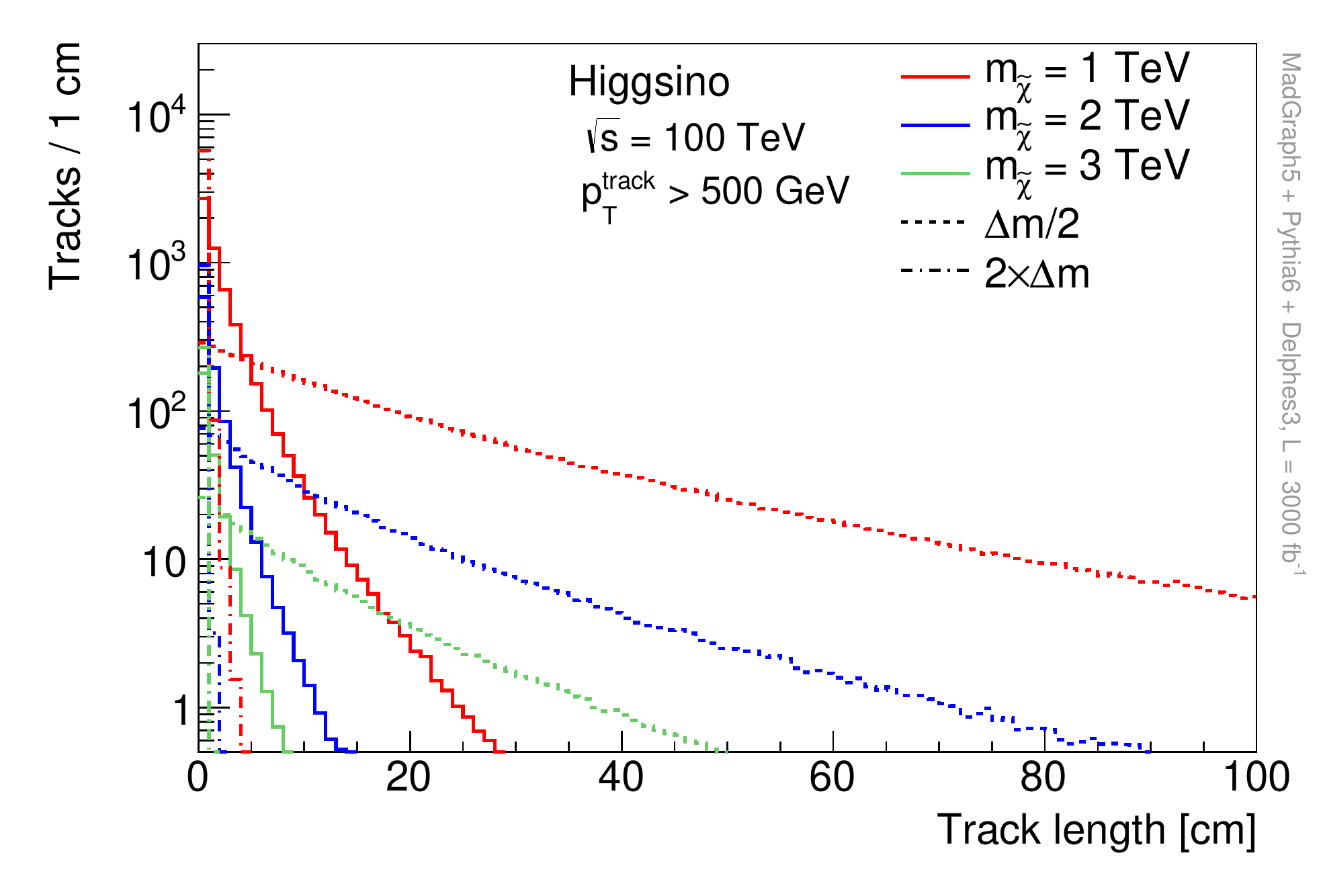} \includegraphics[scale=0.35]{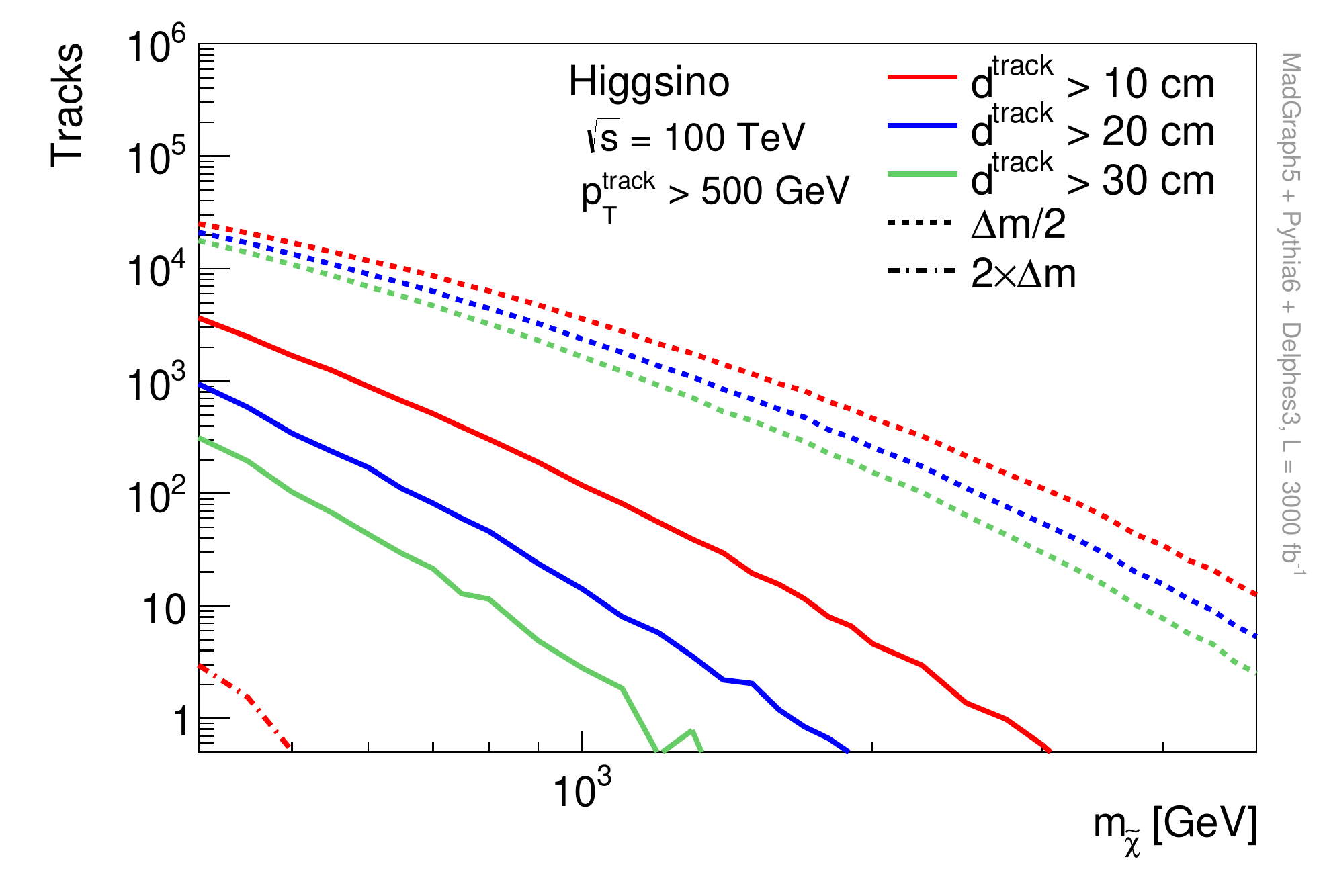}
  \caption{Chargino track distributions for the pure higgsino scenario showing the number of tracks for a given track length (left) and the number of tracks for a given higgsino mass (right).  The dashed lines shows the same plots with a neutralino-chargino mass splitting half the standard value, and the dashed-dotted lines show the same plots with a neutralino-chargino mass splitting twice the standard value.  Only events passing the analysis cuts in App.~\ref{app:analysis} and containing at least one chargino track with $p_T>500~\gev$ are considered.}
  \label{fig:track_distributions_higgsino}
\end{figure}
%%%%%%%%%%%%

Fig.~\ref{fig:monojet_higgsino} shows the mass reach in the monojet channel for the pure higgsino scenario.  As in the wino case, there is a factor 4-5 enhancement in reach for the $100~\tev$ collider relative to the LHC.  The reach is weaker than that for winos, mainly due to the reduction in production cross-section.

Without systematics one finds higgsinos could be excluded at $m_{\tilde{\chi}} \sim 410~\gev$ and discovered at $m_{\tilde{\chi}} \sim 290~\gev$ at 14 TeV, and excluded at $m_{\tilde{\chi}} \sim 1.2~\tev$ and discovered at $m_{\tilde{\chi}} \sim 0.6~\tev$ at 100 TeV.

\para
It is also imaginable to do a disappearing track such for higgsinos.  We note that, in comparison to the wino, it is more likely for heavier new particle states to alter the higgsino splitting as the lowest higher dimensional operator splitting the charged and neutral higgsinos is dimension 5.  Therefore choosing a higgsino splitting has a larger degree of model dependence.  In Fig.~\ref{fig:track_distributions_higgsino} (left) we show the distance of chargino tracks for the standard one-loop splittings, as well as for scenarios with twice the splitting and one half of the splitting.  Fig.~\ref{fig:track_distributions_higgsino} (right) shows the corresponding plot for the number of tracks.

%%%%%%%%%%%%%%%%%%
\begin{figure}[h!]
  \centering
  \includegraphics[scale=0.5]{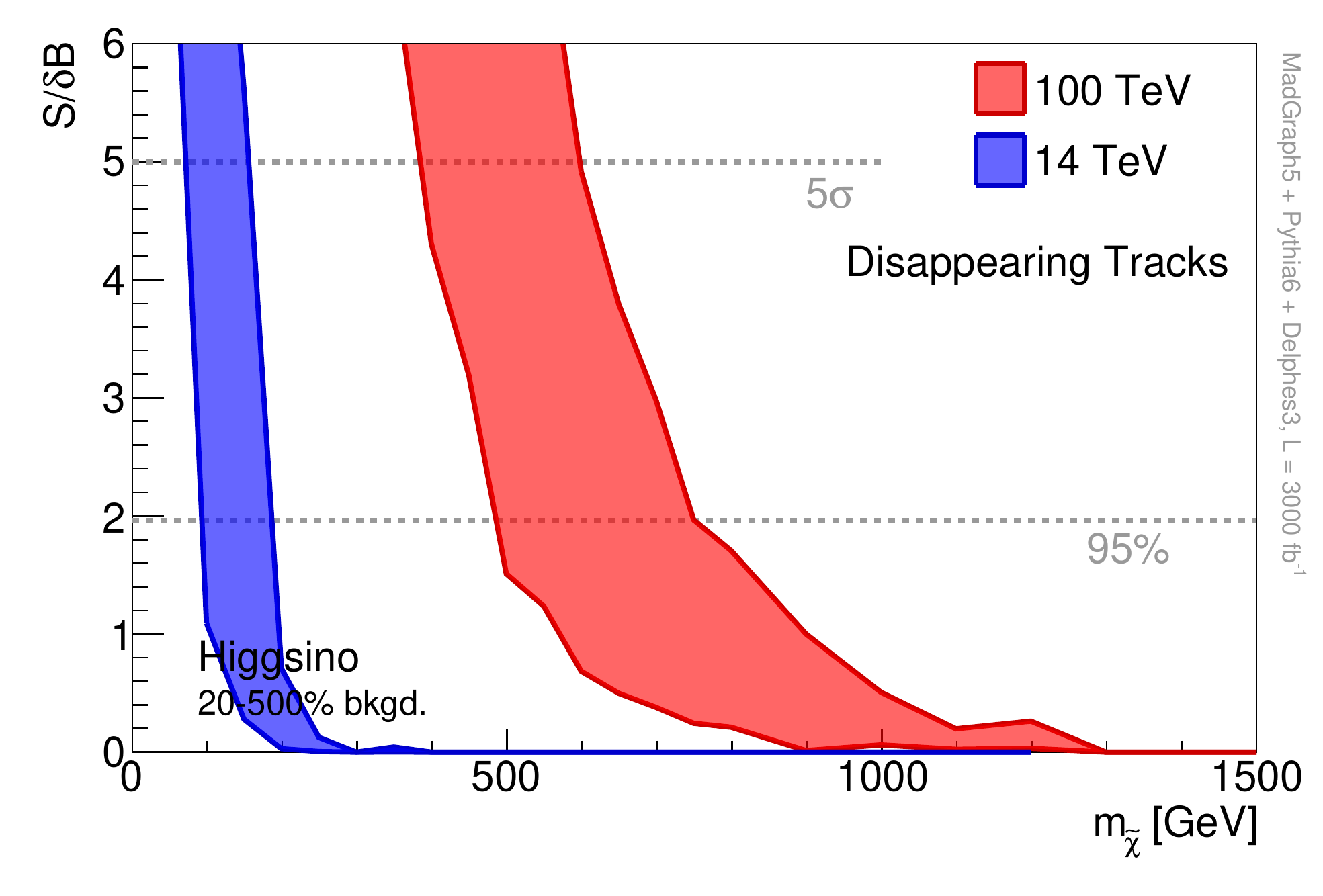}
  \caption{The mass reach in the pure higgsino scenario in the disappearing track channel with $\mathcal{L}=3000~\ifb$ for the $14~\tev$ LHC (blue) and a $100~\tev$ proton-proton collider (red).  The bands are generated by varying the background normalization between $20-500 \%$.  Only events passing the analysis cuts in App.~\ref{app:analysis} are considered.}
  \label{fig:track_higgsino}
\end{figure}
%%%%%%%%%%%%

%%%%%%%%%%%%%%%%%%
\begin{table}[h!]
  \centering
  \begin{tabular}{|c|c|c|c|c|c|}
    \hline
    \multirow{2}{*}{channel} & systematics/ & \multicolumn{2}{c|}{14 TeV} & \multicolumn{2}{c|}{100 TeV} \\ \cline{3-6}
                                         & normalization & $95 \%$ limit  & $5\sigma$ discovery & $95 \%$ limit  & $5\sigma$ discovery   \\ 
    \hline \hline
    \multirow{2}{*}{monojet}             & $1\%$    & $185~\gev$     &  $80~\gev$ & $870~\gev$  & $285~\gev$     \\
                                         & $2\%$    &  $95~\gev$     &  $50~\gev$ & $580~\gev$  &  $80~\gev$     \\ \hline
    \multirow{3}{*}{disappearing tracks} & $20\%$   & $185~\gev$     & $155~\gev$ & $750~\gev$  & $595~\gev$      \\
                                         & $100\%$  & $140~\gev$     &  $95~\gev$ & $615~\gev$  & $485~\gev$      \\ 
                                         & $500\%$  &  $90~\gev$     &  $70~\gev$ & $485~\gev$  & $380~\gev$      \\ \hline
  \end{tabular}
  \caption{Mass reach for the pure higgsino scenario.  For the monojet channel, the second column shows the systematic uncertainty on the background used, while the systematic uncertainty on the signal was $10 \%$.  For the disappearing tracks channel, the second column shows the background normalization.  For this channel the background systematic uncertainty was $20 \%$ and the signal systematic uncertainty was $10 \%$.}
  \label{table:higgsino}
\end{table}
%%%%%%%%%%%%%%%%%%

Results are shown in Table~\ref{table:higgsino}.  We find the monojet channel to reach $m_{\tilde{\chi}} \sim 870~\gev$.  The disappearing track search is potentially a promising channel too, but depends sensitively on the chargino-neutralino mass splitting.  The disappearing track with the canonical splitting is not as sensitive as the monojet search, but were the splitting to be decreased by a factor of two, the limits would be comparable to the reach for winos.

%%%%%%%%%%%%%%%%%%%%%%%%%%%%%%%%%%%%%%%%%%%%%%%%%%%%%%%%%%%%%%%%%%
%%%%%%%%%%%%%%%%%%%%%%%%%%%%%%%%%%%%%%%%%%%%%%%%%%%%%%%%%%%%%%%%%%
\section{Mixed Spectra}
\label{sec:mixed}

%%%%%%%%%%%%%%%%%%
\begin{figure}[h!]
  \centering
  \includegraphics[scale=0.5]{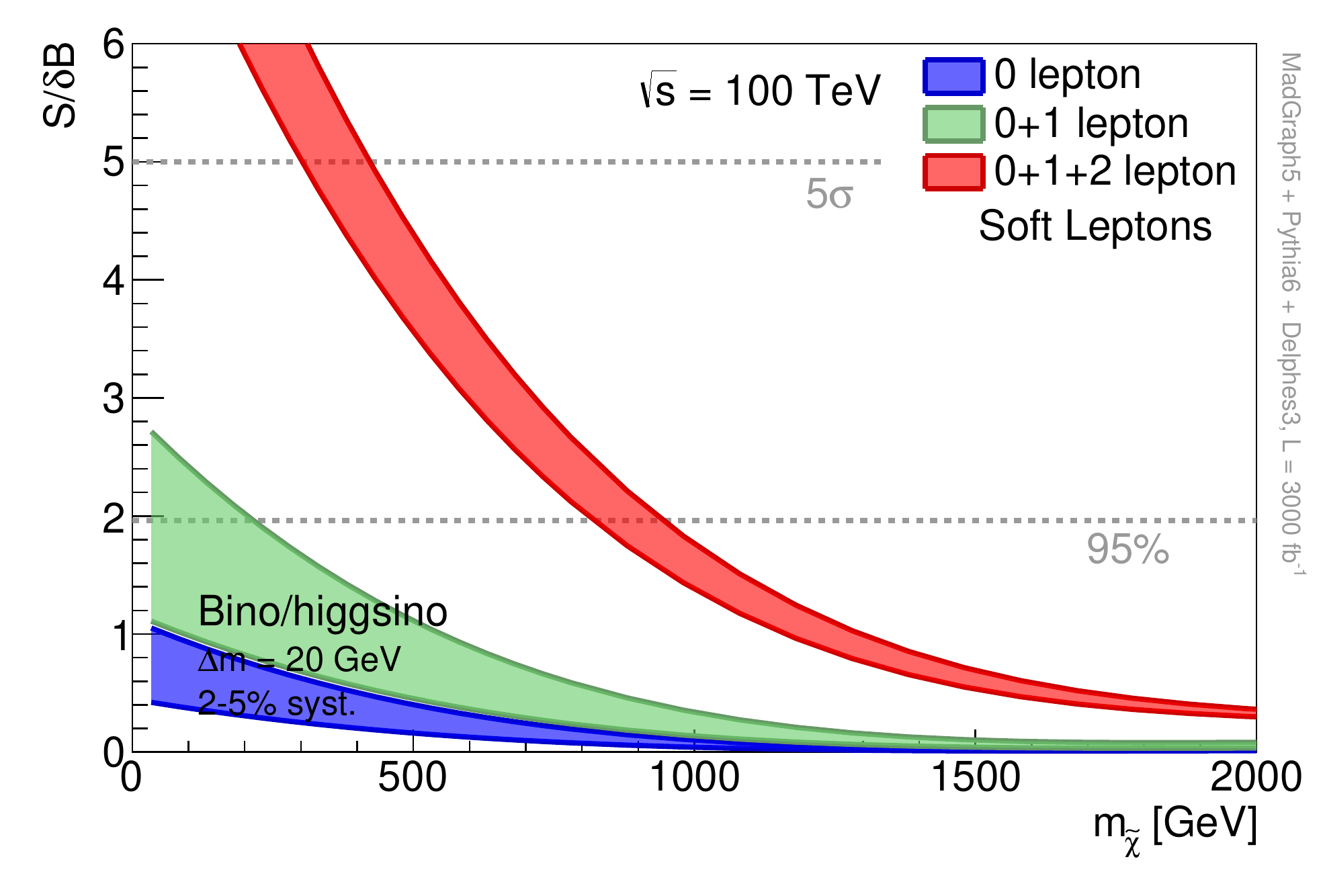}
  \caption{The mass reach in the mixed bino/higgsino ($\Delta = 20~\gev$) scenario in the soft lepton channel at $100~\tev$ with $\mathcal{L}=3000~\ifb$ at $100~\tev$ looking for 0 leptons (blue), 0 or 1 leptons (green), and 0, 1, or 2 leptons (red).  The bands are generated by varying the background systematics between $2-5 \%$ and the signal systematic uncertainty is set to $10 \%$.}
  \label{fig:softlepton_bhtemp20}
\end{figure}
%%%%%%%%%%%%

%%%%%%%%%%%%%%%%%%
\begin{figure}[h!]
  \centering
  \includegraphics[scale=0.5]{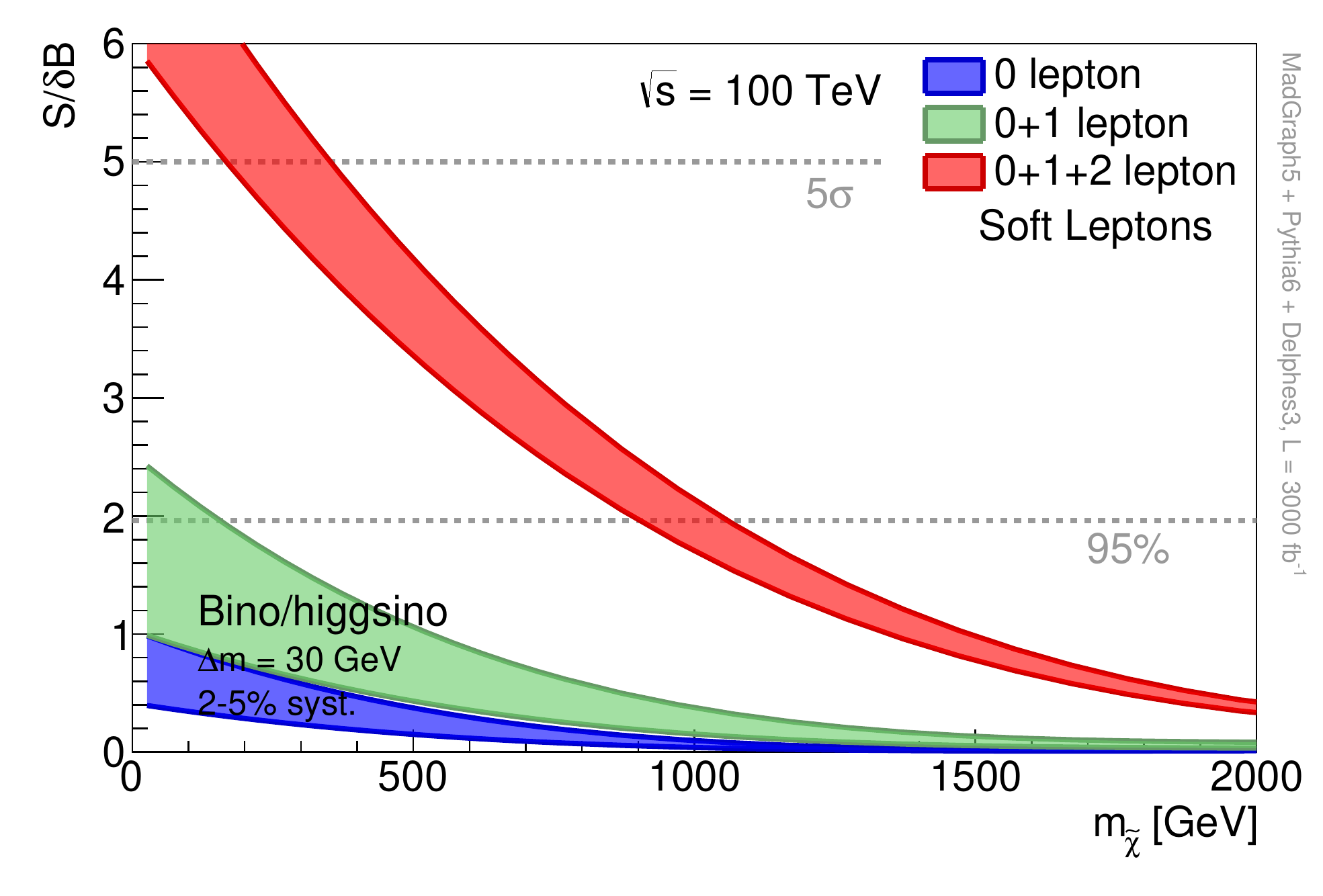}
  \caption{The mass reach in the mixed bino/higgsino ($\Delta = 30~\gev$) scenario in the soft lepton channel at $100~\tev$ with $\mathcal{L}=3000~\ifb$ at $100~\tev$ looking for 0 leptons (blue), 0 or 1 leptons (green), and 0, 1, or 2 leptons (red).  The bands are generated by varying the background systematics between $2-5 \%$ and the signal systematic uncertainty is set to $10 \%$.}
  \label{fig:softlepton_bhtemp30}
\end{figure}
%%%%%%%%%%%%

In the previous two sections we studied the phenomenology of pure LSPs which feature nearly degenerate electroweakinos.  In more general mixed scenarios, larger mass splittings between charginos and neutralinos can be generated.  In this paper, we look at the compressed case of $\Delta m = 20-30~\gev$, where the heavier charginos and neutralinos decay to the LSP via off-shell $W$'s and $Z$'s.

Unlike pure spectra, mixed spectra are known to be able to thermally saturate the relic density, for a range of masses.  Examples include the well-tempered scenario~\cite{ArkaniHamed:2006mb} where $|M_1| \approx |M_2|$ or $|M_1| \approx |\mu|$ and the focus point region~\cite{Chan:1997bi,Feng:1999zg,Feng:2000gh,Feng:2011aa,Akula:2011jx} where the LSP contains a non-trivial higgsino fraction.  To thermally produce the correct relic abundance, one typically needs the LSP is be dominantly higgsino-like or wino-like and subdominantly bino-like.

As we have characterized spectra by their splitting between the LSP and lightest chargino, these spectra typically have very small $\Delta m \gtrsim 0$.  For very small $\Delta m$ the best search tends to be the monojet search (covered in Sects.~\ref{sec:wino},~\ref{sec:higgsino}, and~\ref{sec:coan}).  Because we elect to focus on the collider phenomenology of compressed spectra with soft leptons, we choose spectra with $\Delta m = 20 - 30~\gev$.  While these spectra do not thermally saturate the relic density, they directly demonstrate the utility of soft lepton searches.

Relative to pure winos and higgsinos mixed dark matter can be strongly constrained by direct detection experiments which already and will continue to exclude large regions of parameter space~\cite{Cheung:2012qy,Cahill-Rowley:2013dpa,Cheung:2013dua}.

In this paper we study the three following representative compressed spectra:

\begin{itemize}
  \item ({\it i}) Bino/higgsino $\Delta = 20~\gev$:  We scan over $M_1$ and set $\mu = -M_1 + 23~\gev$.  The low energy states include three neutralinos and a chargino and the mass splitting is $20~\gev$.
  \item ({\it ii}) Bino/higgsino $\Delta = 30~\gev$:  We scan over $M_1$ and set $\mu = -M_1 - 2~\gev$.  The low energy states include three neutralinos and a chargino and the mass splitting is $30~\gev$.
  \item ({\it iii}) Bino/wino(/higgsino) $\Delta = 20~\gev$: We scan over $M_1$ and set $M_2 = M_1 + 34~\gev$ and $\mu = M_1 + 120~\gev$.  The low energy states include all four neutralinos and both charginos and the mass splitting is $20~\gev$.
\end{itemize}

%%%%%%%%%%%%%%%%%%
\begin{figure}[h!]
  \centering
  \includegraphics[scale=0.5]{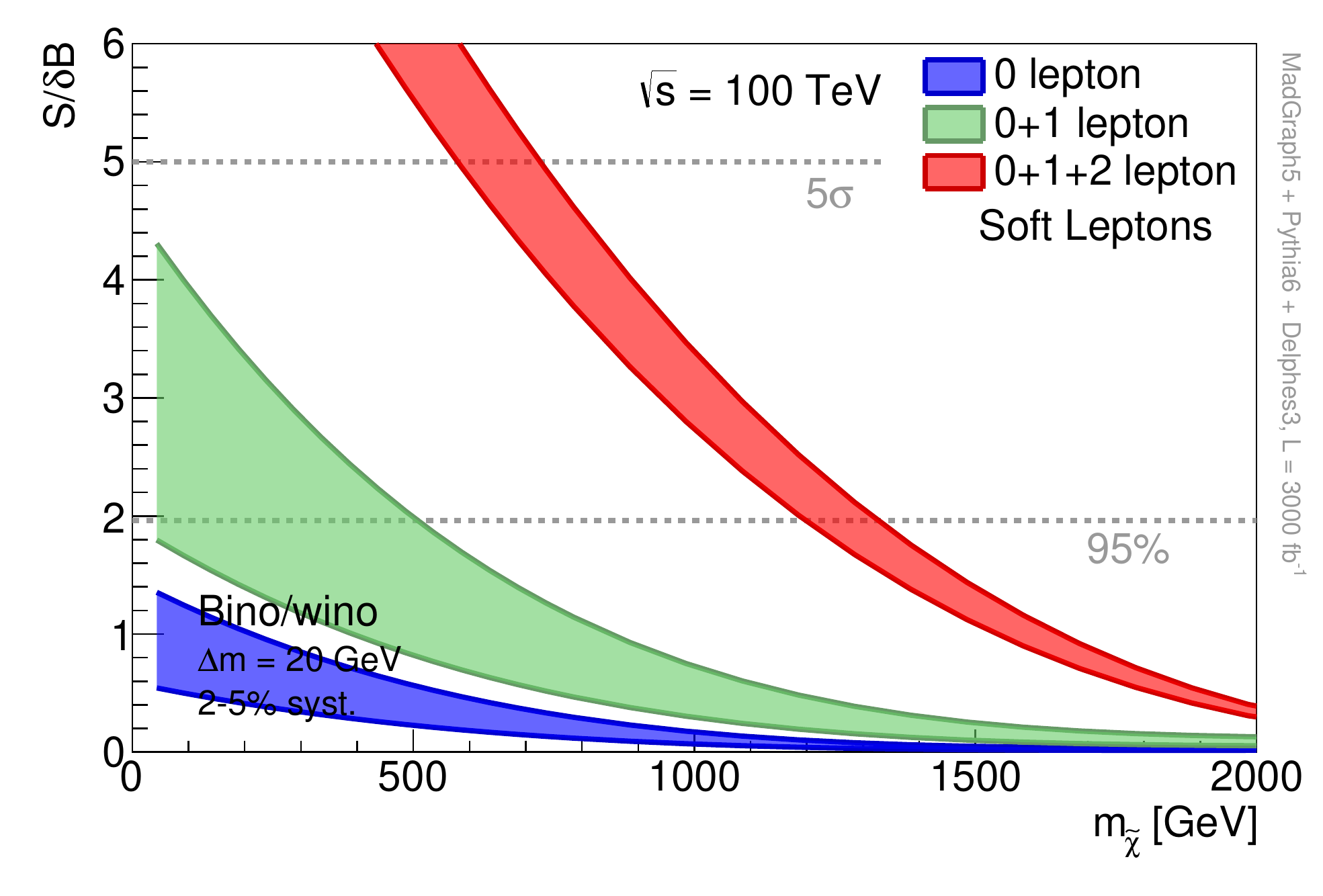}
  \caption{The mass reach in the mixed bino/wino(/higgsino) ($\Delta = 20~\gev$) scenario in the soft lepton channel at $100~\tev$ with $\mathcal{L}=3000~\ifb$ at $100~\tev$ looking for 0 leptons (blue), 0 or 1 leptons (green), and 0, 1, or 2 leptons (red).  The bands are generated by varying the background systematics between $2-5 \%$ and the signal systematic uncertainty is set to $10 \%$.}
  \label{fig:softlepton_allmix20}
\end{figure}
%%%%%%%%%%%%

%%%%%%%%%%%%%%%%%%
\begin{figure}[h!]
  \centering
  \includegraphics[scale=0.5]{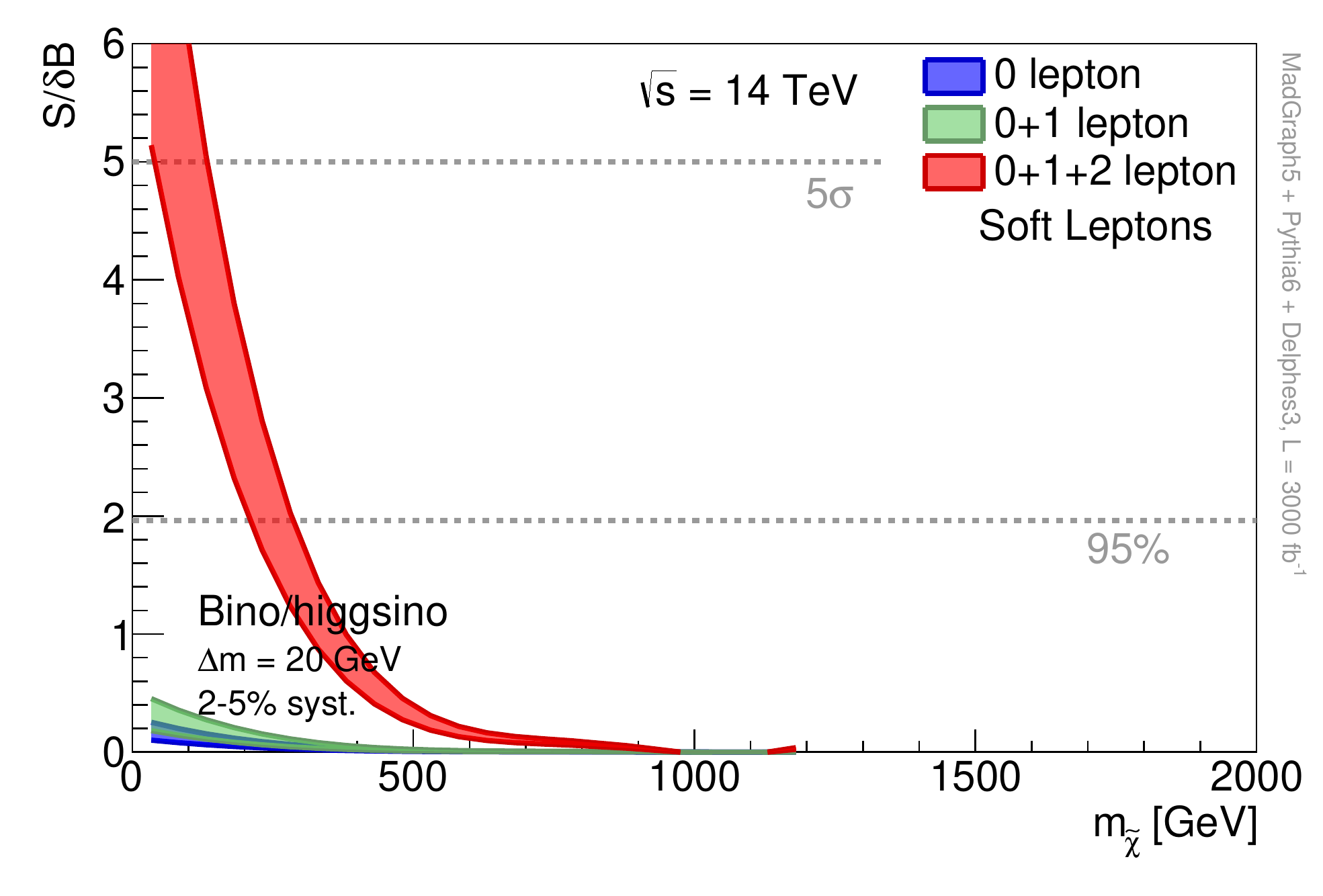}
  \caption{The mass reach in the mixed bino/higgsino ($\Delta = 20~\gev$) scenario in the soft lepton channel at $14~\tev$ with $\mathcal{L}=3000~\ifb$ at $100~\tev$ looking for 0 leptons (blue), 0 or 1 leptons (green), and 0, 1, or 2 leptons (red).  The bands are generated by varying the background systematics between $2-5 \%$ and the signal systematic uncertainty is set to $10 \%$.  The results for bino/higgsino ($\Delta = 30~\gev$) and bino/wino ($\Delta = 20~\gev$) are very similar.}
  \label{fig:softlepton_bhtemp20_ecm14}
\end{figure}
%%%%%%%%%%%%

Figs.~\ref{fig:softlepton_bhtemp20},~\ref{fig:softlepton_bhtemp30}, and~\ref{fig:softlepton_allmix20} show the mass reach for scenarios ({\it i}), ({\it ii}), and ({\it iii}), respectively, in the soft lepton channel at $100~\tev$\footnote{In Figs.~\ref{fig:softlepton_bhtemp20},~\ref{fig:softlepton_bhtemp30},~\ref{fig:softlepton_allmix20}, and~\ref{fig:softlepton_bhtemp20_ecm14} the significances as a function of mass shown are fitted to a 4th order polynomial.  As there are a relatively small number of leptons there are fluctuations in the bands due to statistics.  The fit is not very sensitive to the polynomial used.  Using a 3rd or 5th order polynomial instead only changes the projections by a few GeV.}.  Leptons are considered to be electrons with $10~\gev < p_T < 30~\gev$ or muons with $10~\gev < p_T < 30~\gev$.  In the plots the blue band shows the significance for the 0 lepton bin alone.  Here we set $\lambda = 2 - 5 \%$ due to possibly more sizable systematic uncertainties from identifying low $p_T$ leptons and keep $\gamma = 10 \%$.  The green band shows the significance of the 0 and 1 lepton bins added in quadrature and the red band additionally includes the 2 lepton bin.  Fig.~\ref{fig:softlepton_bhtemp20_ecm14} shows the bino/higgsino scenario for $14~\tev$.  We do not include the corresponding plots for the other mixed spectra as they yield very similar results.

Table~\ref{table:mixed} summarizes our results.  We find in all cases the tagging of soft leptons plays a significant role in maximizing mass reach.  In particular the 2-lepton bin noticeably drives the significance in all cases.  This is because for the 0-lepton and 1-lepton bins the background is dominated by a single boson process.  In the 2-lepton bin, the background is mostly a diboson process, which has a much smaller cross-section.  The exclusion reach extends to $m_{\tilde{\chi}} \sim 1~\tev$ in all cases and the discovery reaches several hundred GeV.

%%%%%%%%%%%%%%%%%%
\begin{table}[t]
  \centering
  \begin{tabular}{|c|c|c|c|c|c|}
    \hline
    \multirow{2}{*}{channel} & 
    \multirow{2}{*}{LSP} & 
    \multirow{2}{*}{$\Delta m$} & 
    \multirow{2}{*}{bkgd. syst.} & 
    \multicolumn{2}{c|}{100 TeV} \\ \cline{5-6}
      &  &  &  & $95 \%$ limit  & $5\sigma$ discovery \\ 
    \hline \hline \multirow{6}{*}{soft leptons} 
    & \multirow{2}{*}{bino/higgsino}& \multirow{2}{*}{$20~\gev$}  & $2\%$ & $940~\gev$ & $420~\gev$  \\
    &                               &                             & $5\%$ & $820~\gev$ & $300~\gev$  \\ \cline{2-6}
    & \multirow{2}{*}{bino/higgsino}& \multirow{2}{*}{$30~\gev$}  & $2\%$ & $1.0~\tev$ & $350~\gev$  \\
    &                               &                             & $5\%$ & $0.9~\tev$ & $165~\gev$  \\ \cline{2-6}
    & \multirow{2}{*}{bino/wino}    & \multirow{2}{*}{$20~\gev$}  & $2\%$ & $1.3~\tev$ & $725~\gev$  \\
    &                               &                             & $5\%$ & $1.2~\tev$ & $575~\gev$  \\ \hline
  \end{tabular}
  \caption{Mass reach for the mixed dark matter scenario.  The systematic uncertainty on the signal was $10 \%$.  At $14~\tev$ the exclusion reach is $m_{\tilde{\chi}} \lesssim 200 - 280~\gev$ (for $5 - 2 \%$) and the discovery reach is $m_{\tilde{\chi}} \lesssim 30 - 130~\gev$ (for $5 - 2 \%$) for all three specta.}
  \label{table:mixed}
\end{table}
%%%%%%%%%%%%%%%%%%

%%%%%%%%%%%%%%%%%%%%%%%%%%%%%%%%%%%%%%%%%%%%%%%%%%%%%%%%%%%%%%%%%%
%%%%%%%%%%%%%%%%%%%%%%%%%%%%%%%%%%%%%%%%%%%%%%%%%%%%%%%%%%%%%%%%%%
\section{Coannihilating Spectra}
\label{sec:coan}

In the discussion of thermally viable dark matter models another relevant set of models are those that include coannihilation.  While bino LSPs alone oversaturate the relic abundance, if another particle is almost mass degenerate it can enhance the annihilation cross-section enough to produce the observed relic density~\cite{Griest:1990kh}.  In this section we explore four different spectra classified by the particles close in energy to the LSP: ({\it i}) gluinos, ({\it ii}) stops, ({\it iii}) squarks, and ({\it iv}) staus.  While coannihilation with wino and higgsino LSPs are also possible, here we only consider a bino LSP.

\para
The first spectrum we consider is gluino coannihilation.  While such spectra do not arise in minimal supergravity, they can be realized in more general non-universal models.  The phenomenology has been studied~\cite{Profumo:2004wk,Feldman:2009zc} and the relic abundance has recently been calculated including Sommerfeld effects~\cite{Harigaya:2014dwa,deSimone:2014pda}\footnote{There is a small difference in the mass splitting values found, $\Delta m \lesssim 10~\gev$, between \cite{Harigaya:2014dwa} and \cite{deSimone:2014pda} for gluino coannihilation.  This difference is due to different choices in renormalization scales.  In what follows we cite values calculated in \cite{deSimone:2014pda}.}.  In the limit of the bino and gluino being mass degenerate, they find $m_{\tilde{\chi}} \sim 7.5~\tev$ produces the correct relic density.

We set $m_{\tilde{g}} - m_{\tilde{\chi}} \approx 0.05 m_{\tilde{\chi}}$ and decouple everything else, leaving one neutralino and the gluinos at low energies.  Relic abundance calculations are very sensitive to the exact mass splitting used but collider processes much less sensitive.  The signal consists of pair produced gluinos, which then decay to the LSP and other standard model particles which could be tagged.  As the decays depend on details of the other SUSY particles we remain agnostic and assume the gluinos decay as $\tilde{g} \to \tilde{\chi}^0_1 + \text{undetected}$.

%%%%%%%%%%%%%%%%%%
\begin{figure}[h!]
  \centering
  \includegraphics[scale=0.5]{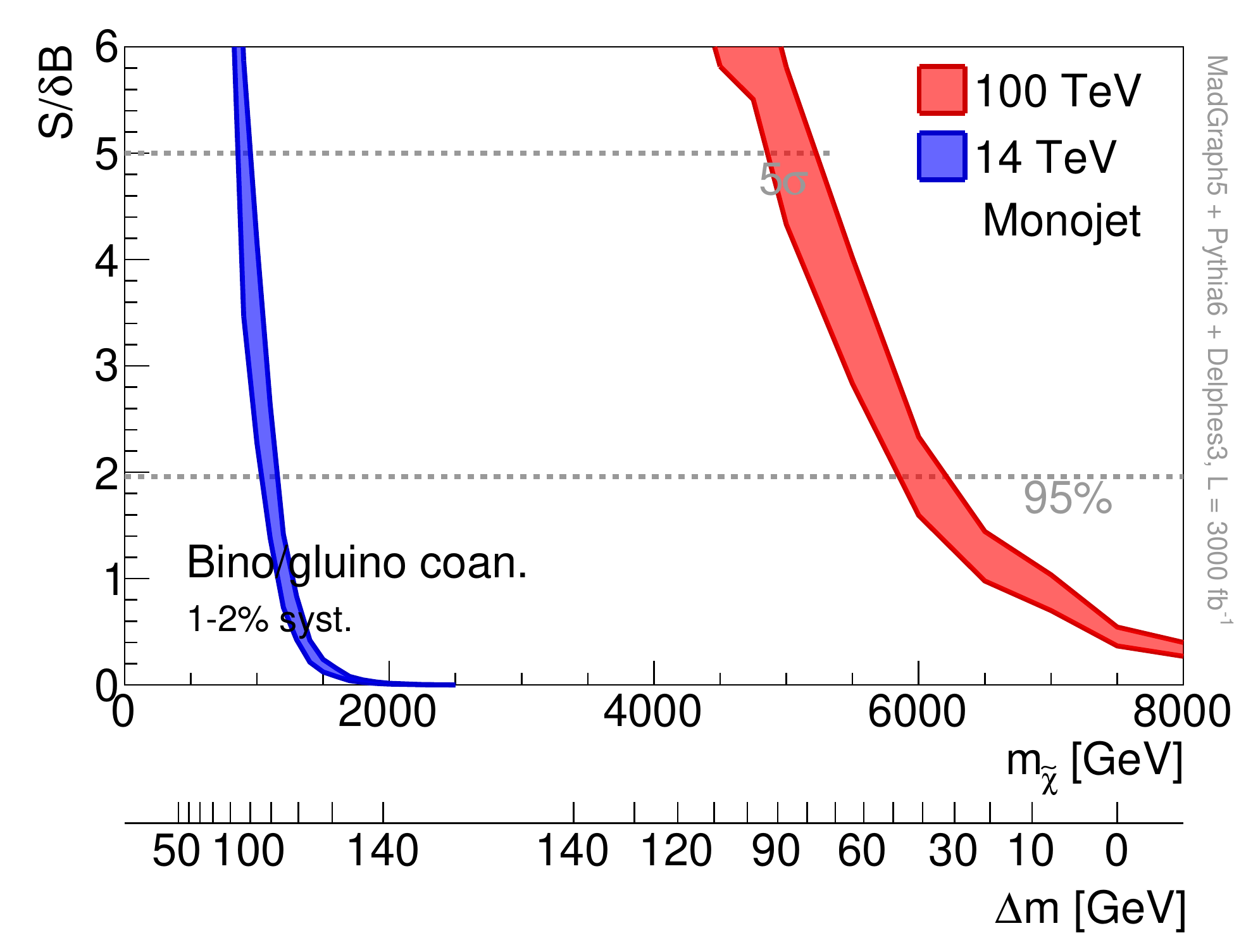}
  \caption{The mass reach in the gluino coannihilation scenario in the monojet channel with $\mathcal{L}=3000~\ifb$ for the $14~\tev$ LHC (blue) and a $100~\tev$ proton-proton collider (red).  The bands are generated by varying the background systematics between $1-2 \%$ and the signal systematic uncertainty is set to $10 \%$.  The lower $x$-axis displays the gluino-bino mass splitting $\Delta m$ for a given bino mass which is required to saturate the relic density~\cite{Harigaya:2014dwa,deSimone:2014pda}.  A tick is placed every $10~\gev$ with the exception of the consecutive $\Delta m=140~\gev$ ticks.}
  \label{fig:monojet_gluino}
\end{figure}
%%%%%%%%%%%%

Fig.~\ref{fig:monojet_gluino} shows the mass reach applying the same monojet search as for the pure wino and higgsino.  In this section we will continue to use $\gamma = 10 \%$ and $\lambda = 1-2 \%$.  Given a bino mass and demanding that it saturate the relic density, one can find a gluino mass for which this is true.  This defines the mass difference $\Delta m = m_{\tilde{g}} - m_{\tilde{\chi}}$ between the gluino and bino and has been calculated as a function of $m_{\tilde{\chi}}$ in~\cite{Harigaya:2014dwa,deSimone:2014pda}.  In Fig.~\ref{fig:monojet_gluino}, the $x$-axis shows the bino mass, while the lower $x$-axis denotes to the gluino-bino splitting required to saturate the relic density for the corresponding bino mass.  As can be seen, a $100~\tev$ monojet search can rule out a bino LSP with $m_{\tilde{\chi}} \sim 6.2~\tev$ and a gluino at $m_{\tilde{g}} \sim 6.23~\tev$ ({\it i.e.} $\Delta m \sim 30~\gev$), but cannot exclude the case where $\Delta m=0$ (corresponding to $m_{\tilde{\chi}} = m_{\tilde{g}} \sim 7.5~\tev$).  If systematics are not considered the mass reach increases by $\sim 350~\gev$ at 14 TeV and by $\sim 250~\gev$ at 100 TeV.

\para
The next coannihilator considered is the stop.  As the mass of the stop is tied to fine-tuning, stop coannihilation appears in many models~\cite{Boehm:1999bj,Cohen:2013kna} and has also been previously studied~\cite{Carena:2008mj}.  In our simulations we set $m_{\tilde{t}} - m_{\tilde{\chi}} \approx 0.05 m_{\tilde{\chi}}$ and decouple everything else, leaving one neutralino and the right-handed stop at low energies.  

The mass reach is shown in Fig.~\ref{fig:monojet_stop}.  The mass for a thermal bino is $m_{\tilde{\chi}} \sim 1.8~\tev$ in the stop-degenerate limit.  A $100~\tev$ collider can not only comfortably exclude this scenario, but also discover it, given sufficiently low systematics.  Without systematics the mass reach increases by $\sim 250~\gev$ at 14 TeV and by $\sim 300~\gev$ at 100 TeV.

%%%%%%%%%%%%%%%%%%
\begin{figure}[h!]
  \centering
  \includegraphics[scale=0.5]{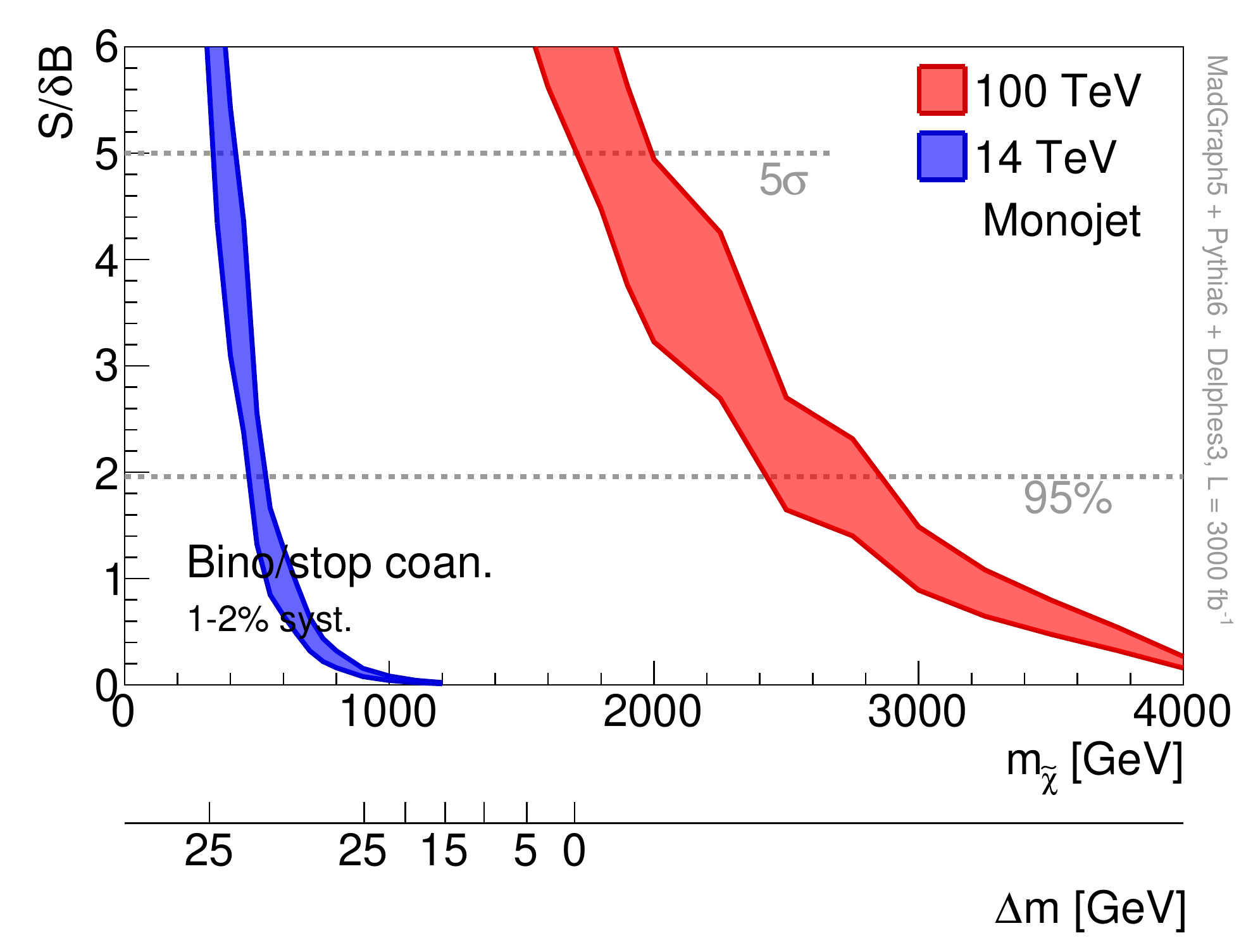}
  \caption{The mass reach in the stop coannihilation scenario in the monojet channel with $\mathcal{L}=3000~\ifb$ for the $14~\tev$ LHC (blue) and a $100~\tev$ proton-proton collider (red).  The bands are generated by varying the background systematics between $1-2 \%$ and the signal systematic uncertainty is set to $10 \%$.  The lower $x$-axis displays the stop-bino mass splitting $\Delta m$ for a given bino mass which is required to satisfy the relic density~\cite{deSimone:2014pda}.  A tick is placed every $5~\gev$ with the exception of the consecutive $\Delta m=25~\gev$ ticks.}
  \label{fig:monojet_stop}
\end{figure}
%%%%%%%%%%%%

\para
Next we move onto squark coannihilation.  For this spectrum we keep the left-handed scalar partners of the light quarks ($\tilde{u}_L$, $\tilde{d}_L$, $\tilde{s}_L$, and $\tilde{c}_L$) in the spectrum, while decoupling everything else.  We set these in the same manner as in the other coannihilation spectra, $m_{\tilde{q}} - m_{\tilde{\chi}} \approx 0.05 m_{\tilde{\chi}}$.  Fig.~\ref{fig:monojet_squark} shows the monojet reach.  As expected the significance is roughly four times larger than the stop coannihilation case.  The exclusion reach extends to $m_{\tilde{\chi}} \sim 4~\tev$ and the discovery reach to $m_{\tilde{\chi}} \sim 3~\tev$.  The mass reach goes up by $\sim 300~\gev$ at 14 TeV and by $\sim 500~\gev$ at 100 TeV if systematics are ignored.

%%%%%%%%%%%%%%%%%%
\begin{figure}[h!]
  \centering
  \includegraphics[scale=0.5]{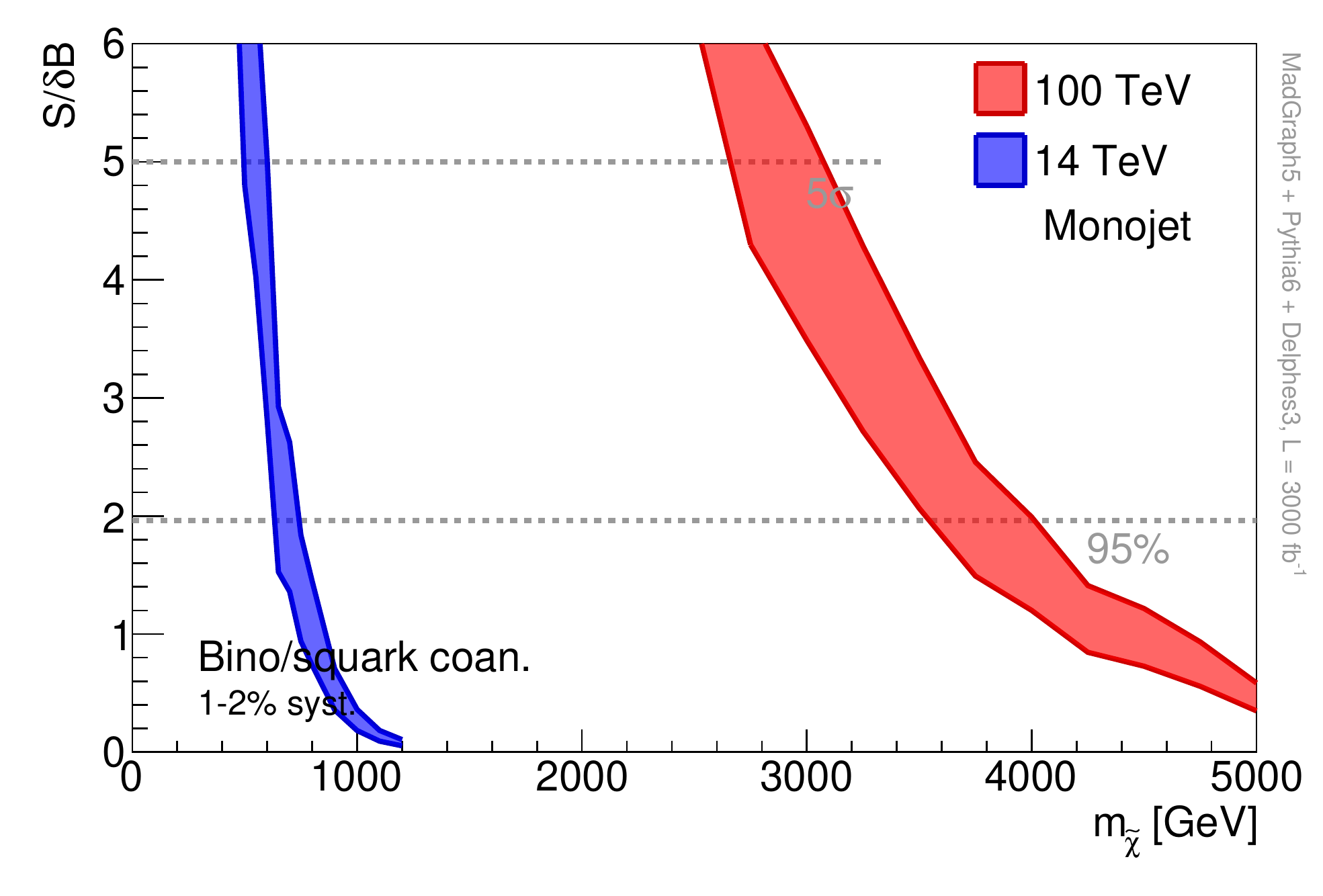}
  \caption{The mass reach in the squark coannihilation scenario in the monojet channel with $\mathcal{L}=3000~\ifb$ for the $14~\tev$ LHC (blue) and a $100~\tev$ proton-proton collider (red).  The bands are generated by varying the background systematics between $1-2 \%$ and the signal systematic uncertainty is set to $10 \%$.}
  \label{fig:monojet_squark}
\end{figure}
%%%%%%%%%%%%

\para
Lastly we studied the stau coannihilation scenario.  These regions come up in constrained MSSM parameter scans, albeit with other particles at low energies~\cite{Ellis:1998kh}.  Again, we set $m_{\tilde{\tau}} - m_{\tilde{\chi}} \approx 0.05 m_{\tilde{\chi}}$, leaving a neutralino and the right-handed stau at low energies.  The cross-section for stau pair production is suppressed by more than an order of magnitude relative to the strongly-interacting coannihilators and is too low for the monojet channel to have any sensitivity~\cite{Lindert:2011td}.  Projecting the reach in constrained MSSM stau coannihilation regions at $100~\tev$ would require a more detailed study involving other particles in the spectrum and different search channels.

%%%%%%%%%%%%%%%%%%
\begin{table}[t]
  \centering
  \begin{tabular}{|c|c|c|c|c|c|c|}
    \hline
    \multirow{2}{*}{channel} & 
    \multirow{2}{*}{coannihilator} & 
    \multirow{2}{*}{bkgd. syst.} & 
    \multicolumn{2}{c|}{14 TeV} & 
    \multicolumn{2}{c|}{100 TeV} \\ \cline{4-7}
      &  &  & $95 \%$ limit  & $5\sigma$ discovery & $95 \%$ limit  & $5\sigma$ discovery \\ 
    \hline \hline \multirow{8}{*}{monojet} 
    & \multirow{2}{*}{gluino} & $1\%$ & $1.1~\tev$ & $950~\gev$ & $6.2~\tev$     & $5.2~\tev$ \\
    &                         & $2\%$ & $1.0~\tev$ & $850~\gev$ & $5.8~\tev$     & $4.8~\tev$ \\ \cline{2-7}
    & \multirow{2}{*}{stop}   & $1\%$ & $530~\gev$ & $420~\gev$ & $2.8~\tev$     & $2.1~\tev$ \\
    &                         & $2\%$ & $470~\gev$ & $330~\gev$ & $2.4~\tev$     & $1.7~\tev$ \\ \cline{2-7}
    & \multirow{2}{*}{squark} & $1\%$ & $740~\gev$ & $600~\gev$ & $4.0~\tev$     & $3.0~\tev$ \\
    &                         & $2\%$ & $630~\gev$ & $495~\gev$ & $3.5~\tev$     & $2.6~\tev$ \\ \cline{2-7}
    & \multirow{2}{*}{stau}   & \multirow{2}{*}{n/a} & \multirow{2}{*}{n/a} & \multirow{2}{*}{n/a} & \multirow{2}{*}{n/a} & \multirow{2}{*}{n/a} \\
    &                         &                      &                      &                      &                      &                      \\ \hline
  \end{tabular}
  \caption{Mass reach for the coannihilating dark matter scenario.  The systematic uncertainty on the signal was $10 \%$.}
  \label{table:coan}
\end{table}
%%%%%%%%%%%%%%%%%%

\para
The coannihilation results are summarized in Table~\ref{table:coan}.  To recapitulate we find the exclusion reach for gluinos to be $m_{\tilde{\chi}} \sim 6.2~\tev$, for stops to be $m_{\tilde{\chi}} \sim 2.8~\tev$, and for squarks to be $m_{\tilde{\chi}} \sim 4.0~\tev$.  The monojet search is not sensitive to the stau coannihilation scenario.  The discovery prospects are also all in the multi-TeV range.

%%%%%%%%%%%%%%%%%%%%%%%%%%%%%%%%%%%%%%%%%%%%%%%%%%%%%%%%%%%%%%%%%%
%%%%%%%%%%%%%%%%%%%%%%%%%%%%%%%%%%%%%%%%%%%%%%%%%%%%%%%%%%%%%%%%%%
\section{Conclusions}
\label{sec:conclusions}

In this work we projected the $95 \%$ exclusion reach and the $5 \sigma$ discovery reach of a $100~\tev$ proton-proton collider for neutralino dark matter.  As SUSY already provides a variety of basic dark matter models we performed our study in the context of simplified SUSY models, but the results can be straightfowardly generalized.  We implemented three collider searches, all of which relied on the basic signal of tagging one or more initial state radiation (ISR) jets and summarized in Fig.~\ref{fig:summary}.

\para
The first spectrum studied was pure wino dark matter.  Recently, wino dark matter has received some attention based on the potential to exclude or discover it with indirect detection experiments.  Unfortunately, the LHC is only able to probe the several hundred $\gev$ range, which is neither near the thermally-saturating wino mass, nor high enough to close the available mass window from the low end (given a pessimistic dark matter halo profile).  A $100~\tev$ collider, in contrast, can exclude as high as $m_{\tilde{\chi}} \sim 1.4~\tev$ in the monojet channel, or even $m_{\tilde{\chi}} \sim 3~\tev$ given a naive extrapolation of a disappearing track search.  In light both of $8~\tev$ LHC results and this study, it is clear that the disappearing track (and displaced vertices and charged massive particle) search will play an exigent role in continuing to carve away at wino parameter space.

\para
Higgsino dark matter was next to be looked at and was found to receive similar enhancements in mass reach in going from the LHC to a $100~\tev$ collider as wino dark matter.  The higgsino cross-section, however, is lower than that of the wino, which is reflected in a lower exclusion and discovery reach.  Respectively these reaches were found to be $m_{\tilde{\chi}} \sim 870~\gev$ and $m_{\tilde{\chi}} \sim 285~\gev$.  The chargino-neutralino mass splitting for higgsinos is parametrically larger than the wino mass splitting leading to short chargino track length and a less sensitive disappearing track search.  The monojet or disappearing track searches alone are not likely to quite reach the thermal higgsino mass.  One direction of future work would be to examine combining several searches to reach the thermal higgsino mass or augment the spectrum with additional particles to open up new search channels.

%%%%%%%%%%%%%%%%%%
\begin{figure}[h!]
  \centering
  \includegraphics[scale=0.75]{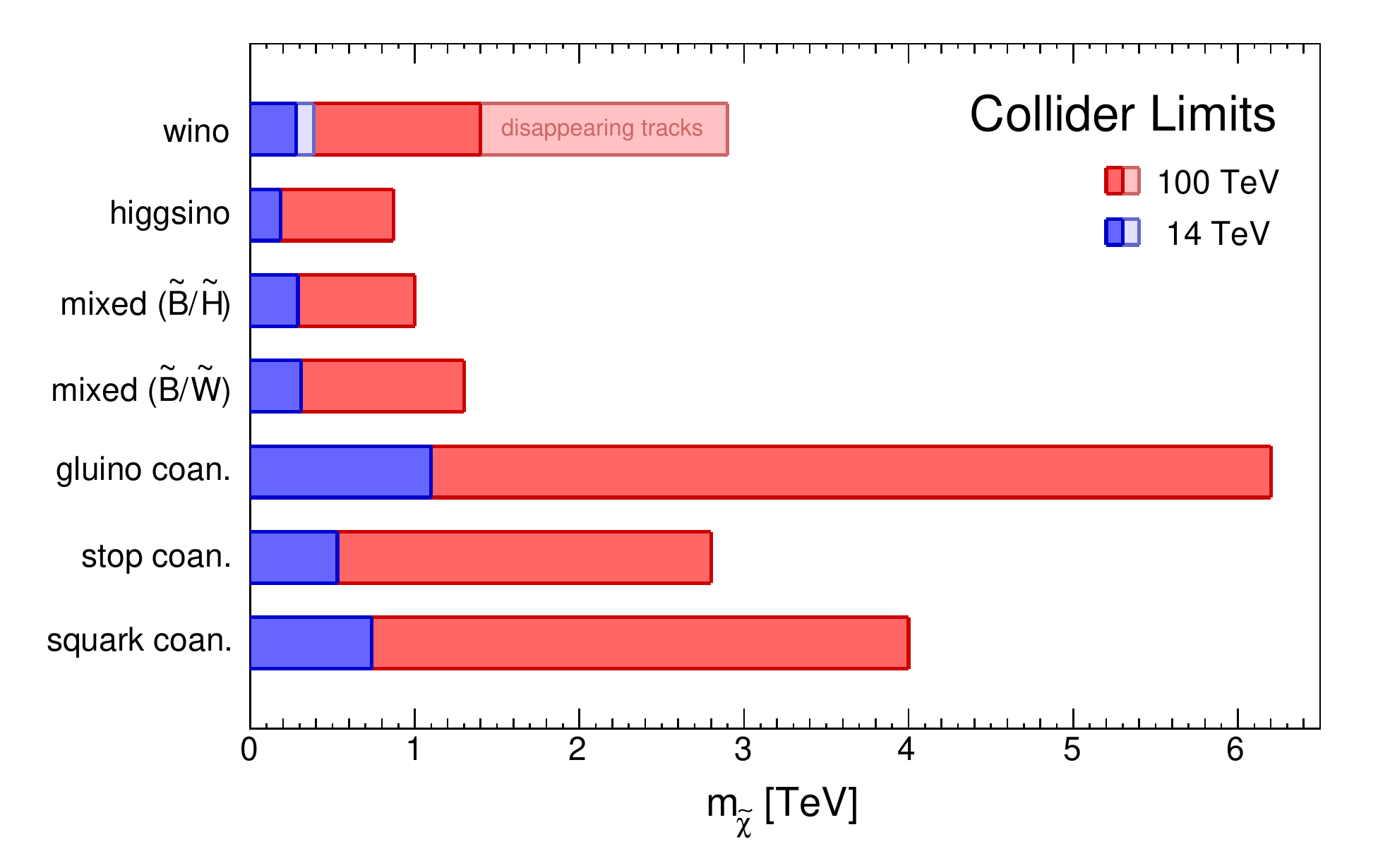}
  \caption{Summary of collider reach for neutralino dark matter.}
  \label{fig:summary}
\end{figure}
%%%%%%%%%%%%

\para
The next spectra were several cases of mixed dark matter with a compressed spectrum of $\Delta m = 20-30~\gev$.  In these cases the most applicable search was looking for soft leptons in association with the hard ISR jet(s).  The exclusion reach was found to be $m_{\tilde{\chi}} \sim 1~\tev$, while the discovery reach ranged from $350-700~\gev$.  Mixed dark matter parameter space already receives strong constraints from direct detection and a more thorough study on the impact of collider searches on this parameter space would be worthwhile.

\para
Finally bino dark matter was studied, bringing various coannihilators into the spectrum to avoid overclosing the universe.  These scenarios utilized the monojet search to project reach.  The stop coannihilation exclusion reach was found to be $m_{\tilde{\chi}} \sim 2.8~\tev$ and the discovery reach to be $m_{\tilde{\chi}} \sim 2.1~\tev$.  As the thermally-saturating bino mass in this case is $m_{\tilde{\chi}} \sim 1.8~\tev$ (and $m_{\tilde{t}} \sim 1.8~\tev$), dark matter can be either excluded or discovered in this channel.  The gluino coannihilation, on the other hand, was found to only reach the thermal bino mass for a splitting of $\Delta m = 30~\gev$, corresponding to $m_{\tilde{\chi}} \sim 6.2~\tev$ and $m_{\tilde{g}} \sim 6.23~\tev$, so the thermal parameter space is not entirely closed.  Finally squark coannihilation can be excluded up to $m_{\tilde{\chi}} \sim 4.0~\tev$ and stau coannihilation cannot be probed in the monojet channel.

\para
In addition to the aforementioned interplay with mixed dark matter and neutralino blindspots, useful future work would be to look at how adding in more search channels can improve the dark matter collider reach.  Such searches would include monophoton searches, razor searches, vector boson fusions searches, and multilepton searches.  Another principal direction to extend these studies would be to look at the impact of bringing down other particles into the low energy spectrum.

\para
In conclusion, we have shown that while the $14~\tev$ LHC can probe dark matter of several hundred $\gev$, a $100~\tev$ proton-proton collider could foray directly into the multi-TeV range.  As this is precisely the range where thermally-saturating neutralino dark matter lives, a $100~\tev$ collider has great potential for discovering WIMP dark matter.  Were a discovery to be made in the near future, a $100~\tev$ collider will be a vital tool for understanding the nature of dark matter.

%%%%%%%%%%%%%%%%%%%%%%%%%%%%%%%%%

\acknowledgments

The authors would like to thank Nima Arkani-Hamed, Alan Barr, Philip Harris, Shigeki Matsumoto, Steven Schramm, Pedro Schwaller, and Alessandro Strumia, as well as the participants of BSM Opportunities at 100 TeV Workshop for useful discussions.

ML is supported by NSERC of Canada and LTW is supported by DOE grant DE-SC0003930.  Computations for this paper were performed on the Hypnotoad cluster supported by PSD Computing at the University of Chicago.

%%%%%%%%%%%%%%%%%%%%%%%%%%%%%%%%%%%%%%%%%%%%%%%%%%%%%%%%%%%%%%%%%%
%%%%%%%%%%%%%%%%%%%%%%%%%%%%%%%%%%%%%%%%%%%%%%%%%%%%%%%%%%%%%%%%%%
\appendix
\section{Analysis Details}
\label{app:analysis}

\subsection*{Monojet}

Our monojet analysis closely follows the $8~\tev$ CMS monojet analysis~\cite{CMS:2013rwa}.  To validate our background monte carlo and analysis we duplicated the CMS monojet cuts at $8~\tev$ and found good agreement across the cut flow.  The only discrepancy present in our simulation was a slightly increased lepton efficiency due to the settings of the Snowmass detector card we used.

The $8~\tev$ analysis proceeds as:

\begin{itemize}
  \item Require a hard central jet $p_T(j_1)>110~\gev$, $|\eta(j_1)|<2.4$.
  \item A second jet, defined as $p_T(j_2)>30~\gev$, $|\eta(j_2)|<4.5$, is permitted, but any additional jets are vetoed $n_{\text{jet}}\leq2$.
  \item If there is a second jet it cannot be back-to-back with the hardest jet $\Delta\phi(j_1,j_2)<2.5$.
  \item Identified leptons are vetoed.  To be identified, electrons need to have $p_T(e) > 10~\gev$ and $|\eta(e)|<2.5$, muons need to have $p_T(\mu)>10~\gev$ and $|\eta(\mu)|<2.1$ and reconstructed hadronic taus need to have $p_T(\tau) > 20~\gev$ and $|\eta(\tau)|<2.3$.
  \item The signal regions are defined in overlapping $\met$ bins: $>250~\gev$, $>300~\gev$, $>350~\gev$, $>400~\gev$, $>450~\gev$, $>500~\gev$, and $>550~\gev$.  The bin with the highest expected significance is used to set the limit.
\end{itemize}

At $14~\tev$ and $100~\tev$ the cuts were adjusted as shown in Tables~\ref{table:monojet_cuts} and~\ref{table:monojet_met}.

%%%%%%%%%%%%%%%%%%
\begin{table}[t]
  \centering
  \begin{tabular}{|c|c|c|c|} \hline
    Cut                       & $8~\tev$          & $14~\tev$            & $100~\tev$            \\ \hline \hline
    $p_T(j_1)$, $\eta(j_1)$   & $110~\gev$, $2.4$ & $300~\gev$, $2.4$    & $1200~\gev$, $2.4$    \\
    $p_T(j_2)$, $\eta(j_2)$   & $30~\gev$, $4.5$  & $30-120~\gev$, $4.5$ & $100-400~\gev$, $4.5$ \\
    $n_{\text{jet}}$          & $2$               & $2$                  & $2$                   \\
    $\Delta\phi(j_1,j_2)$     & $2.5$             & $2.5$                & $2.5$                 \\
    $p_T(e)$, $\eta(e)$       & $10~\gev$, $2.5$  & $20~\gev$, $2.5$     & $20~\gev$, $2.5$      \\
    $p_T(\mu)$, $\eta(\mu)$   & $10~\gev$, $2.1$  & $20~\gev$, $2.1$     & $20~\gev$, $2.1$      \\
    $p_T(\tau)$, $\eta(\tau)$ & $20~\gev$, $2.3$  & $30~\gev$, $2.3$     & $40~\gev$, $2.3$      \\
    $\met$                    & $250-550~\gev$    & $350-1000~\gev$      & $2-5~\tev$            \\
    \hline
  \end{tabular}
  \caption{Cuts used in monojet analysis.  For $p_T(j_2)$ and $\met$ the range represents the values scanned over, where the values used for each spectra are shown in Table~\ref{table:monojet_met}.}
  \label{table:monojet_cuts}
\end{table}
%%%%%%%%%%%%%%%%%%

\para
In light of the higher rates and luminosity for the $14~\tev$ and $100~\tev$ we also explored if the significance can be improved by applying additional cuts.  We found that allowing for a third jet and making cuts on the various distances between jets did not help because the signal and backgrounds have very similar kinematics.  The distributions are not identical, however, and it is possible that a clever analysis, like razor~\cite{Rogan:2010kb,Fox:2012ee,Buckley:2013kua} for instance, could still yield a small improvement.

%%%%%%%%%%%%%%%%%%
\begin{table}[t]
  \centering
  \begin{tabular}{|c|c|c c c c c c|} \hline
   $\sqrt{s}$                   & Cut        & Wino       & Higgsino   & Gluino coan. & Stop coan. & Squark coan. & Stau coan.  \\ 
   \hline \hline
   \multirow{2}{*}{$14~\tev$}   & $\met$     & $650~\gev$ & $650~\gev$ & $750~\gev$   & $650~\gev$ & $650~\gev$ & $650~\gev$   \\ 
                                & $p_T(j_2)$ & $30~\gev$  & $30~\gev$  & $120~\gev$   & $120~\gev$ & $120~\gev$ & $120~\gev$   \\ \hline
   \multirow{2}{*}{$100~\tev$}  & $\met$     & $3.5~\tev$ & $3.5~\tev$ & $4.0~\tev$   & $3.5~\tev$ & $3.5~\tev$ & $3.5~\tev$   \\ 
                                & $p_T(j_2)$ & $300~\gev$ & $250~\gev$ & $400~\gev$   & $400~\gev$ & $400~\gev$ & $400~\gev$   \\ 
   \hline
  \end{tabular}
  \caption{$\met$ and $p_T(j_2)$ cuts used in the monojet analysis for each spectra.  Table~\ref{table:monojet_cuts} shows the other cuts used.}
  \label{table:monojet_met}
\end{table}
%%%%%%%%%%%%%%%%%%

% add comment that stau values don't matter

\subsection*{Soft Leptons}

The soft lepton analysis is similar to the monojet analysis with the exception of how leptons are treated.  Events are binned according to whether they contain $0$, $1$, or $2$ leptons where a lepton is defined as an electron with $10~\gev < p_T(e) < 30~\gev$ or a muon with $10~\gev < p_T(\mu) < 30~\gev$.  Hadronic taus are not tagged.  The bins are assumed to be uncorrelated and their significances are added in quadtrature.

%%%%%%%%%%%%%%%%%%
\begin{table}[t]
  \centering
  \begin{tabular}{|c|c|c|} \hline
    Cut                        & $100~\tev$                      & $14~\tev$                           \\ \hline \hline
    $p_T(j_1)$, $\eta(j_1)$    & $1200~\gev$, $2.4$              & $300~\gev$, $2.4$                   \\
    $p_T(j_2)$, $\eta(j_2)$    & $300~\gev$, $4.5$               & $30~\gev$, $4.5$                    \\
    $n_{\text{jet}}$           & $2$                             & $2$                                 \\
    $\Delta\phi(j_1,j_2)$      & $2.5$                           & $2.5$                               \\
    $p_T(e)$, $\eta(e)$        & $\in (10~\gev,30~\gev)$, $2.5$  & $\in (10~\gev,30~\gev)$, $2.5$      \\
    $p_T(\mu)$, $\eta(\mu)$    & $\in (10~\gev,30~\gev)$, $2.1$  & $\in (10~\gev,30~\gev)$, $2.1$      \\
    $\met$                     & $1250~\gev$                     & $350~\gev$                          \\
    \hline
  \end{tabular}
  \caption{Cuts used in soft lepton analysis.}
  \label{table:softlep_cuts}
\end{table}
%%%%%%%%%%%%%%%%%%

\subsection*{Disappearing Tracks}

For signal events we replicate the analysis in the $8~\tev$ ATLAS analysis~\cite{Aad:2013yna}, which applies the following cuts

\begin{itemize}
  \item Require a hard jet $p_T(j_1) > 90~\gev$.
  \item Require that $\met > 90~\gev$.
  \item If there are any other jets with $p_T(j_2) > 45~\gev$, the hardest of these is considered the second jet.
  \item Compute the azimuthal separation, $\Delta\phi(j,\met)$, between the missing energy and the hardest jet.  If there is a second jet and its azimuthal separation from the missing energy is smaller, use that instead.  Only keep events where $\Delta\phi_{\text{min}}(j,\met) > 1.5$.
  \item There must be at least chargino track that is isolated and satisfies a track selection criteria and $0.1<|\eta^{\text{track}}|<1.9$.
  \item Signal regions are defined by a $p_T$ cut on the chargino track.  The bins are $p_T^{\text{track}}>75~\gev$, $p_T^{\text{track}}>100~\gev$, $p_T^{\text{track}}>150~\gev$, and $p_T^{\text{track}}>200~\gev$.
\end{itemize}

We implement the isolation cut by rejecting events with jets within $\Delta R < 0.4$ of the chargino track, where $\Delta R = \sqrt{ (\Delta \eta)^2 + (\Delta \phi)^2}$.  To mock up the good track selection we assume the efficiency factors as $\epsilon_{\text{track}} = \epsilon_{\text{det}} \times \epsilon_{\text{tracker}}$, where we assume $\epsilon_{\text{tracker}}$ is $100 \%$ for tracks with a length $30~\cm < d^{\text{track}} < 80~\gev$ and $0 \%$ otherwise and that $\epsilon_{\text{det}}$ is flat with respect to $p_T^{\text{track}}$ and $\eta^{\text{track}}$.  We derive $\epsilon_{\text{det}}$ by matching our event count from monte carlo to the event count from~\cite{Aad:2013yna}.

%%%%%%%%%%%%%%%%%%
\begin{table}[t]
  \centering
  \begin{tabular}{|c|c|c|c|} \hline
    Cut                                & $8~\tev$        & $14~\tev$       & $100~\tev$      \\ \hline \hline
    $\met$                             & $90~\gev$       & $130~\gev$      & $975~\gev$      \\
    $p_T(j_1)$                         & $90~\gev$       & $130~\gev$      & $975~\gev$      \\
    $p_T(j_2)$                         & $45~\gev$       & $70~\gev$       & $500~\gev$      \\
    $\Delta\phi_{\text{min}}(j,\met)$  & $1.5$           & $1.5$           & $1.5$           \\
    $\eta^{\text{track}}$              & $\in (0.1,1.9)$ & $\in (0.1,1.9)$ & $\in (0.1,1.9)$ \\
    $p_T^{\text{track}}$               & $75-200~\gev$   & $250~\gev$      & $1.5~\tev$      \\
    \hline
  \end{tabular}
  \caption{Cuts used in disappearing track analysis.}
  \label{table:track_cuts}
\end{table}
%%%%%%%%%%%%%%%%%%

In the ATLAS $8~\tev$ study, in the signal region the dominant background is from mismeasured tracks and found to fit to a power law

\begin{equation}
  \frac{d\sigma}{dp_T} = \sigma_0 p_T^{-a} ,
\end{equation}
where $p_T$ is measured in $\gev$, $\sigma_0$ is the normalization, and $a=1.78 \pm 0.05$.  We extrapolate this by assuming that the majority of events passing the cuts in Table~\ref{table:track_cuts} are $Z(\nu\nu)+\text{jets}$ events.  Under this assumption we scale the normalization according to the $Z(\nu\nu)+\text{jets}$ cross-section after the cuts in Table~\ref{table:track_cuts} and make the appropriate $p_T$ cut.  We find $\sigma_0$ by matching to the ATLAS result.

%%%%%%%%%%%%%%%%%%
\section{Extrapolating Systematics}
\label{app:systematics}

While it is impossible to project the systematic uncertainties for a not-yet-designed detector, many of the systematics are generic features of detector technology.  For completeness we list the dominant contributions to the various analyses here.  More detailed discussions are found in the appropriate experimental analyses~\cite{ATLAS:2012ky,Chatrchyan:2012me,ATLAS:2012zim,CMS:2013rwa,Aad:2013yna}.

The dominant contributions to systematics are
\begin{itemize}

  \item {\it Jet uncertainties}: the jet energy scale and jet energy resolution determine how well jets, and consequently $\slashed{E}_T$, can be measured.  These reduce at higher luminosity as the detector becomes better understood.  Calorimeter measurements should be better at higher energies so these will likely naturally decrease at a 100 TeV collider (assuming a deep enough calorimeter).

  \item {\it Monte carlo uncertainties}: backgrounds are typically computed by counting events in a control region and extrapolating to the signal region using ratios computed by monte carlo generators.  The uncertainty on the backgrounds is determined by the number of events in the control regions and the monte carlo extrapolation.  At high luminosity the control regions will contain more events reducing the statistical uncertainties.  This will also help create control regions for subdominant backgrounds, {\it e.g.} $t\bar{t}$ and diboson, which are currently determined entirely from monte carlo.  These backgrounds can have uncertainties as large as $\sim 100 \%$.

 \para
  Additionally, monte carlo modeling can be improved by including higher-order perturbative corrections.  Given the recent advancements in monte carlo technology (see~\cite{Salam:2011bj} and references therein) it may be reasonable to expect similar leaps by the time a 100 TeV collider is built.

  \item {\it Lepton uncertainties}: precise measurements of lepton momenta are not only important in the soft lepton analysis, but also in the monojet and disappearing track analyses.  While leptons are vetoed in the latter two searches, they are important in defining the control regions.  Unfortunately muons are detected via tracking which degrades at higher $p_T$.  It will be important for the detectors at 100 TeV to maintain the ability to measure energetic muons to a reasonable precision.

\end{itemize}

\bibliographystyle{jhep}
\bibliography{darkmatter}

\providecommand{\href}[2]{#2}\begingroup\raggedright\begin{thebibliography}{100}

\bibitem{Bertone:2004pz}
G.~Bertone, D.~Hooper, and J.~Silk, {\it {Particle dark matter: Evidence,
  candidates and constraints}},  {\em Phys.Rept.} {\bf 405} (2005) 279--390,
  [\href{http://xxx.lanl.gov/abs/hep-ph/0404175}{{\tt hep-ph/0404175}}].

\bibitem{Jungman:1995df}
G.~Jungman, M.~Kamionkowski, and K.~Griest, {\it {Supersymmetric dark matter}},
   {\em Phys.Rept.} {\bf 267} (1996) 195--373,
  [\href{http://xxx.lanl.gov/abs/hep-ph/9506380}{{\tt hep-ph/9506380}}].

\bibitem{Martin:1997ns}
S.~P. Martin, {\it {A Supersymmetry primer}},
  \href{http://xxx.lanl.gov/abs/hep-ph/9709356}{{\tt hep-ph/9709356}}.

\bibitem{Alves:2011wf}
LHC New Physics Working Group, D.~Alves et~al., {\it {Simplified Models for LHC
  New Physics Searches}},  {\em J.Phys.} {\bf G39} (2012) 105005,
  [\href{http://xxx.lanl.gov/abs/1105.2838}{{\tt arXiv:1105.2838}}].

\bibitem{Cohen:2013xda}
T.~Cohen, T.~Golling, M.~Hance, A.~Henrichs, K.~Howe, et~al., {\it {SUSY
  Simplified Models at 14, 33, and 100 TeV Proton Colliders}},
  \href{http://xxx.lanl.gov/abs/1311.6480}{{\tt arXiv:1311.6480}}.

\bibitem{Fowlie:2014awa}
A.~Fowlie and M.~Raidal, {\it {Prospects for constrained supersymmetry at
  $\sqrt{s}=33$ TeV and $\sqrt{s}=100$ TeV proton-proton super-colliders}},
  \href{http://xxx.lanl.gov/abs/1402.5419}{{\tt arXiv:1402.5419}}.

\bibitem{Wells:2003tf}
J.~D. Wells, {\it {Implications of supersymmetry breaking with a little
  hierarchy between gauginos and scalars}},
  \href{http://xxx.lanl.gov/abs/hep-ph/0306127}{{\tt hep-ph/0306127}}.

\bibitem{ArkaniHamed:2004fb}
N.~Arkani-Hamed and S.~Dimopoulos, {\it {Supersymmetric unification without low
  energy supersymmetry and signatures for fine-tuning at the LHC}},  {\em JHEP}
  {\bf 0506} (2005) 073, [\href{http://xxx.lanl.gov/abs/hep-th/0405159}{{\tt
  hep-th/0405159}}].

\bibitem{Giudice:2004tc}
G.~Giudice and A.~Romanino, {\it {Split supersymmetry}},  {\em Nucl.Phys.} {\bf
  B699} (2004) 65--89, [\href{http://xxx.lanl.gov/abs/hep-ph/0406088}{{\tt
  hep-ph/0406088}}].

\bibitem{Beltran:2010ww}
M.~Beltran, D.~Hooper, E.~W. Kolb, Z.~A. Krusberg, and T.~M. Tait, {\it
  {Maverick dark matter at colliders}},  {\em JHEP} {\bf 1009} (2010) 037,
  [\href{http://xxx.lanl.gov/abs/1002.4137}{{\tt arXiv:1002.4137}}].

\bibitem{Fox:2011pm}
P.~J. Fox, R.~Harnik, J.~Kopp, and Y.~Tsai, {\it {Missing Energy Signatures of
  Dark Matter at the LHC}},  {\em Phys.Rev.} {\bf D85} (2012) 056011,
  [\href{http://xxx.lanl.gov/abs/1109.4398}{{\tt arXiv:1109.4398}}].

\bibitem{Gershtein:2008bf}
Y.~Gershtein, F.~Petriello, S.~Quackenbush, and K.~M. Zurek, {\it {Discovering
  hidden sectors with mono-photon $Z^\prime$o searches}},  {\em Phys.Rev.} {\bf
  D78} (2008) 095002, [\href{http://xxx.lanl.gov/abs/0809.2849}{{\tt
  arXiv:0809.2849}}].

\bibitem{Fox:2011fx}
P.~J. Fox, R.~Harnik, J.~Kopp, and Y.~Tsai, {\it {LEP Shines Light on Dark
  Matter}},  {\em Phys.Rev.} {\bf D84} (2011) 014028,
  [\href{http://xxx.lanl.gov/abs/1103.0240}{{\tt arXiv:1103.0240}}].

\bibitem{Bai:2012xg}
Y.~Bai and T.~M. Tait, {\it {Searches with Mono-Leptons}},  {\em Phys.Lett.}
  {\bf B723} (2013) 384--387, [\href{http://xxx.lanl.gov/abs/1208.4361}{{\tt
  arXiv:1208.4361}}].

\bibitem{Petriello:2008pu}
F.~J. Petriello, S.~Quackenbush, and K.~M. Zurek, {\it {The Invisible
  $Z^\prime$ at the CERN LHC}},  {\em Phys.Rev.} {\bf D77} (2008) 115020,
  [\href{http://xxx.lanl.gov/abs/0803.4005}{{\tt arXiv:0803.4005}}].

\bibitem{Carpenter:2012rg}
L.~M. Carpenter, A.~Nelson, C.~Shimmin, T.~M. Tait, and D.~Whiteson, {\it
  {Collider searches for dark matter in events with a Z boson and missing
  energy}},  {\em Phys.Rev.} {\bf D87} (2013), no.~7 074005,
  [\href{http://xxx.lanl.gov/abs/1212.3352}{{\tt arXiv:1212.3352}}].

\bibitem{Petrov:2013nia}
A.~A. Petrov and W.~Shepherd, {\it {Searching for dark matter at LHC with
  Mono-Higgs production}},  {\em Phys.Lett.} {\bf B730} (2014) 178--183,
  [\href{http://xxx.lanl.gov/abs/1311.1511}{{\tt arXiv:1311.1511}}].

\bibitem{Carpenter:2013xra}
L.~Carpenter, A.~DiFranzo, M.~Mulhearn, C.~Shimmin, S.~Tulin, et~al., {\it
  {Mono-Higgs: a new collider probe of dark matter}},
  \href{http://xxx.lanl.gov/abs/1312.2592}{{\tt arXiv:1312.2592}}.

\bibitem{Berlin:2014cfa}
A.~Berlin, T.~Lin, and L.-T. Wang, {\it {Mono-Higgs Detection of Dark Matter at
  the LHC}},  \href{http://xxx.lanl.gov/abs/1402.7074}{{\tt arXiv:1402.7074}}.

\bibitem{Schwaller:2013baa}
P.~Schwaller and J.~Zurita, {\it {Compressed electroweakino spectra at the
  LHC}},  \href{http://xxx.lanl.gov/abs/1312.7350}{{\tt arXiv:1312.7350}}.

\bibitem{Han:2013usa}
C.~Han, A.~Kobakhidze, N.~Liu, A.~Saavedra, L.~Wu, et~al., {\it {Probing Light
  Higgsinos in Natural SUSY from Monojet Signals at the LHC}},  {\em JHEP} {\bf
  1402} (2014) 049, [\href{http://xxx.lanl.gov/abs/1310.4274}{{\tt
  arXiv:1310.4274}}].

\bibitem{Bhattacherjee:2013wna}
B.~Bhattacherjee, A.~Choudhury, K.~Ghosh, and S.~Poddar, {\it {Compressed SUSY
  at 14 TeV LHC}},  {\em Phys.Rev.} {\bf D89} (2014) 037702,
  [\href{http://xxx.lanl.gov/abs/1308.1526}{{\tt arXiv:1308.1526}}].

\bibitem{Baer:2014cua}
H.~Baer, A.~Mustafayev, and X.~Tata, {\it {Monojets and mono-photons from light
  higgsino pair production at LHC14}},
  \href{http://xxx.lanl.gov/abs/1401.1162}{{\tt arXiv:1401.1162}}.

\bibitem{Han:2014kaa}
Z.~Han, G.~D. Kribs, A.~Martin, and A.~Menon, {\it {Hunting Quasi-Degenerate
  Higgsinos}},  \href{http://xxx.lanl.gov/abs/1401.1235}{{\tt
  arXiv:1401.1235}}.

\bibitem{Zhou:2013raa}
N.~Zhou, D.~Berge, L.~Wang, D.~Whiteson, and T.~Tait, {\it {Sensitivity of
  future collider facilities to WIMP pair production via effective operators
  and light mediators}},  \href{http://xxx.lanl.gov/abs/1307.5327}{{\tt
  arXiv:1307.5327}}.

\bibitem{Djouadi:2002ze}
A.~Djouadi, J.-L. Kneur, and G.~Moultaka, {\it {SuSpect: A Fortran code for the
  supersymmetric and Higgs particle spectrum in the MSSM}},  {\em
  Comput.Phys.Commun.} {\bf 176} (2007) 426--455,
  [\href{http://xxx.lanl.gov/abs/hep-ph/0211331}{{\tt hep-ph/0211331}}].

\bibitem{Alwall:2011uj}
J.~Alwall, M.~Herquet, F.~Maltoni, O.~Mattelaer, and T.~Stelzer, {\it {MadGraph
  5 : Going Beyond}},  {\em JHEP} {\bf 1106} (2011) 128,
  [\href{http://xxx.lanl.gov/abs/1106.0522}{{\tt arXiv:1106.0522}}].

\bibitem{Sjostrand:2006za}
T.~Sjostrand, S.~Mrenna, and P.~Z. Skands, {\it {PYTHIA 6.4 Physics and
  Manual}},  {\em JHEP} {\bf 0605} (2006) 026,
  [\href{http://xxx.lanl.gov/abs/hep-ph/0603175}{{\tt hep-ph/0603175}}].

\bibitem{deFavereau:2013fsa}
DELPHES 3, J.~de~Favereau et~al., {\it {DELPHES 3, A modular framework for fast
  simulation of a generic collider experiment}},  {\em JHEP} {\bf 1402} (2014)
  057, [\href{http://xxx.lanl.gov/abs/1307.6346}{{\tt arXiv:1307.6346}}].

\bibitem{Anderson:2013kxz}
J.~Anderson, A.~Avetisyan, R.~Brock, S.~Chekanov, T.~Cohen, et~al., {\it
  {Snowmass Energy Frontier Simulations}},
  \href{http://xxx.lanl.gov/abs/1309.1057}{{\tt arXiv:1309.1057}}.

\bibitem{Avetisyan:2013dta}
A.~Avetisyan, S.~Bhattacharya, M.~Narain, S.~Padhi, J.~Hirschauer, et~al., {\it
  {Snowmass Energy Frontier Simulations using the Open Science Grid (A Snowmass
  2013 whitepaper)}},  \href{http://xxx.lanl.gov/abs/1308.0843}{{\tt
  arXiv:1308.0843}}.

\bibitem{Avetisyan:2013onh}
A.~Avetisyan, J.~M. Campbell, T.~Cohen, N.~Dhingra, J.~Hirschauer, et~al., {\it
  {Methods and Results for Standard Model Event Generation at $\sqrt{s}$ = 14
  TeV, 33 TeV and 100 TeV Proton Colliders (A Snowmass Whitepaper)}},
  \href{http://xxx.lanl.gov/abs/1308.1636}{{\tt arXiv:1308.1636}}.

\bibitem{Cacciari:2008gp}
M.~Cacciari, G.~P. Salam, and G.~Soyez, {\it {The Anti-k(t) jet clustering
  algorithm}},  {\em JHEP} {\bf 0804} (2008) 063,
  [\href{http://xxx.lanl.gov/abs/0802.1189}{{\tt arXiv:0802.1189}}].

\bibitem{Cacciari:2011ma}
M.~Cacciari, G.~P. Salam, and G.~Soyez, {\it {FastJet User Manual}},  {\em
  Eur.Phys.J.} {\bf C72} (2012) 1896,
  [\href{http://xxx.lanl.gov/abs/1111.6097}{{\tt arXiv:1111.6097}}].

\bibitem{Beenakker:1999xh}
W.~Beenakker, M.~Klasen, M.~Kramer, T.~Plehn, M.~Spira, et~al., {\it {The
  Production of charginos / neutralinos and sleptons at hadron colliders}},
  {\em Phys.Rev.Lett.} {\bf 83} (1999) 3780--3783,
  [\href{http://xxx.lanl.gov/abs/hep-ph/9906298}{{\tt hep-ph/9906298}}].

\bibitem{Cullen:2012eh}
G.~Cullen, N.~Greiner, and G.~Heinrich, {\it {Susy-QCD corrections to
  neutralino pair production in association with a jet}},  {\em Eur.Phys.J.}
  {\bf C73} (2013) 2388, [\href{http://xxx.lanl.gov/abs/1212.5154}{{\tt
  arXiv:1212.5154}}].

\bibitem{Abazov:2003gp}
D0 Collaboration, V.~Abazov et~al., {\it {Search for large extra dimensions in
  the monojet + missing $E_T$ channel at D\O}},  {\em Phys.Rev.Lett.} {\bf 90}
  (2003) 251802, [\href{http://xxx.lanl.gov/abs/hep-ex/0302014}{{\tt
  hep-ex/0302014}}].

\bibitem{Aaltonen:2008hh}
CDF Collaboration, T.~Aaltonen et~al., {\it {Search for large extra dimensions
  in final states containing one photon or jet and large missing transverse
  energy produced in $p \bar{p}$ collisions at $\sqrt{s}$ = 1.96-TeV}},  {\em
  Phys.Rev.Lett.} {\bf 101} (2008) 181602,
  [\href{http://xxx.lanl.gov/abs/0807.3132}{{\tt arXiv:0807.3132}}].

\bibitem{Abazov:2008kp}
D0 Collaboration, V.~Abazov et~al., {\it {Search for large extra dimensions via
  single photon plus missing energy final states at $\sqrt{s}$ = 1.96-TeV}},
  {\em Phys.Rev.Lett.} {\bf 101} (2008) 011601,
  [\href{http://xxx.lanl.gov/abs/0803.2137}{{\tt arXiv:0803.2137}}].

\bibitem{ATLAS:2012ky}
ATLAS Collaboration, G.~Aad et~al., {\it {Search for dark matter candidates and
  large extra dimensions in events with a jet and missing transverse momentum
  with the ATLAS detector}},  {\em JHEP} {\bf 1304} (2013) 075,
  [\href{http://xxx.lanl.gov/abs/1210.4491}{{\tt arXiv:1210.4491}}].

\bibitem{Chatrchyan:2012me}
CMS Collaboration, S.~Chatrchyan et~al., {\it {Search for dark matter and large
  extra dimensions in monojet events in $pp$ collisions at $\sqrt{s}=7$ TeV}},
  {\em JHEP} {\bf 1209} (2012) 094,
  [\href{http://xxx.lanl.gov/abs/1206.5663}{{\tt arXiv:1206.5663}}].

\bibitem{ATLAS:2012zim}
ATLAS Collaboration, {\it {Search for New Phenomena in Monojet plus Missing
  Transverse Momentum Final States using 10fb-1 of pp Collisions at sqrt{s}=8
  TeV with the ATLAS detector at the LHC}}, ATLAS-CONF-2012-147, ATLAS-COM-CONF-2012-190.

\bibitem{CMS:2013rwa}
CMS Collaboration, {\it {Search for new physics in monojet events in pp
  collisions at sqrt(s)= 8 TeV}}, CMS-PAS-EXO-12-048.

\bibitem{An:2012ue}
H.~An, R.~Huo, and L.-T. Wang, {\it {Searching for Low Mass Dark Portal at the
  LHC}},  {\em Phys.Dark Univ.} {\bf 2} (2013) 50--57,
  [\href{http://xxx.lanl.gov/abs/1212.2221}{{\tt arXiv:1212.2221}}].

\bibitem{An:2012va}
H.~An, X.~Ji, and L.-T. Wang, {\it {Light Dark Matter and $Z'$ Dark Force at
  Colliders}},  {\em JHEP} {\bf 1207} (2012) 182,
  [\href{http://xxx.lanl.gov/abs/1202.2894}{{\tt arXiv:1202.2894}}].

\bibitem{An:2013xka}
H.~An, L.-T. Wang, and H.~Zhang, {\it {Dark matter with $t$-channel mediator: a
  simple step beyond contact interaction}},
  \href{http://xxx.lanl.gov/abs/1308.0592}{{\tt arXiv:1308.0592}}.

\bibitem{CMS:2013dea}
CMS Collaboration, {\it {Search for electroweak production of charginos,
  neutralinos, and sleptons using leptonic final states in pp collisions at 8
  TeV}}, CMS-PAS-SUS-13-006.

\bibitem{CMS:2013jfa}
CMS Collaboration, {\it {A search for anomalous production of events with three
  or more leptons using 19.5/fb of sqrt(s)=8 TeV LHC data}}, CMS-PAS-SUS-13-002.

\bibitem{CMS:2013ida}
CMS Collaboration, {\it {Search for supersymmetry in pp collisions at sqrt(s) =
  8 TeV in events with three leptons and at least one b-tagged jet}}, CMS-PAS-SUS-13-008.

\bibitem{Aad:2014nua}
ATLAS Collaboration, G.~Aad et~al., {\it {Search for direct production of
  charginos and neutralinos in events with three leptons and missing transverse
  momentum in $\sqrt{s}$ = 8 TeV $pp$ collisions with the ATLAS detector}},
  \href{http://xxx.lanl.gov/abs/1402.7029}{{\tt arXiv:1402.7029}}.

\bibitem{Aad:2014vma}
ATLAS Collaboration, G.~Aad et~al., {\it {Search for direct production of
  charginos, neutralinos and sleptons in final states with two leptons and
  missing transverse momentum in pp collisions at $\sqrt{s}$ = 8 TeV with the
  ATLAS detector}},  \href{http://xxx.lanl.gov/abs/1403.5294}{{\tt
  arXiv:1403.5294}}.

\bibitem{Giudice:2010wb}
G.~F. Giudice, T.~Han, K.~Wang, and L.-T. Wang, {\it {Nearly Degenerate
  Gauginos and Dark Matter at the LHC}},  {\em Phys.Rev.} {\bf D81} (2010)
  115011, [\href{http://xxx.lanl.gov/abs/1004.4902}{{\tt arXiv:1004.4902}}].

\bibitem{Gori:2013ala}
S.~Gori, S.~Jung, and L.-T. Wang, {\it {Cornering electroweakinos at the LHC}},
   \href{http://xxx.lanl.gov/abs/1307.5952}{{\tt arXiv:1307.5952}}.

\bibitem{Thomas:1998wy}
S.~D. Thomas and J.~D. Wells, {\it {Phenomenology of Massive Vectorlike Doublet
  Leptons}},  {\em Phys.Rev.Lett.} {\bf 81} (1998) 34--37,
  [\href{http://xxx.lanl.gov/abs/hep-ph/9804359}{{\tt hep-ph/9804359}}].

\bibitem{Cirelli:2005uq}
M.~Cirelli, N.~Fornengo, and A.~Strumia, {\it {Minimal dark matter}},  {\em
  Nucl.Phys.} {\bf B753} (2006) 178--194,
  [\href{http://xxx.lanl.gov/abs/hep-ph/0512090}{{\tt hep-ph/0512090}}].

\bibitem{Feng:1994mq}
J.~L. Feng and M.~J. Strassler, {\it {Determination of fundamental
  supersymmetry parameters from chargino production at LEP-2}},  {\em
  Phys.Rev.} {\bf D51} (1995) 4661--4694,
  [\href{http://xxx.lanl.gov/abs/hep-ph/9408359}{{\tt hep-ph/9408359}}].

\bibitem{Feng:1996ag}
J.~L. Feng and M.~J. Strassler, {\it {Measuring SUSY parameters at LEP-2 using
  chargino production and decay}},  {\em Phys.Rev.} {\bf D55} (1997)
  1326--1342, [\href{http://xxx.lanl.gov/abs/hep-ph/9606477}{{\tt
  hep-ph/9606477}}].

\bibitem{Feng:1999fu}
J.~L. Feng, T.~Moroi, L.~Randall, M.~Strassler, and S.-f. Su, {\it {Discovering
  supersymmetry at the Tevatron in wino LSP scenarios}},  {\em Phys.Rev.Lett.}
  {\bf 83} (1999) 1731--1734,
  [\href{http://xxx.lanl.gov/abs/hep-ph/9904250}{{\tt hep-ph/9904250}}].

\bibitem{Gunion:1999jr}
J.~F. Gunion and S.~Mrenna, {\it {A Study of SUSY signatures at the Tevatron in
  models with near mass degeneracy of the lightest chargino and neutralino}},
  {\em Phys.Rev.} {\bf D62} (2000) 015002,
  [\href{http://xxx.lanl.gov/abs/hep-ph/9906270}{{\tt hep-ph/9906270}}].

\bibitem{Gunion:2001fu}
J.~F. Gunion and S.~Mrenna, {\it {Probing models with near degeneracy of the
  chargino and LSP at a linear e+ e- collider}},  {\em Phys.Rev.} {\bf D64}
  (2001) 075002, [\href{http://xxx.lanl.gov/abs/hep-ph/0103167}{{\tt
  hep-ph/0103167}}].

\bibitem{Barr:2002ex}
A.~Barr, C.~Lester, M.~A. Parker, B.~Allanach, and P.~Richardson, {\it
  {Discovering anomaly mediated supersymmetry at the LHC}},  {\em JHEP} {\bf
  0303} (2003) 045, [\href{http://xxx.lanl.gov/abs/hep-ph/0208214}{{\tt
  hep-ph/0208214}}].

\bibitem{Ibe:2006de}
M.~Ibe, T.~Moroi, and T.~Yanagida, {\it {Possible Signals of Wino LSP at the
  Large Hadron Collider}},  {\em Phys.Lett.} {\bf B644} (2007) 355--360,
  [\href{http://xxx.lanl.gov/abs/hep-ph/0610277}{{\tt hep-ph/0610277}}].

\bibitem{Buckley:2009kv}
M.~R. Buckley, L.~Randall, and B.~Shuve, {\it {LHC Searches for Non-Chiral
  Weakly Charged Multiplets}},  {\em JHEP} {\bf 1105} (2011) 097,
  [\href{http://xxx.lanl.gov/abs/0909.4549}{{\tt arXiv:0909.4549}}].

\bibitem{Aad:2013yna}
ATLAS Collaboration, G.~Aad et~al., {\it {Search for charginos nearly
  mass-degenerate with the lightest neutralino based on a disappearing-track
  signature in pp collisions at $\sqrt{s}$ = 8 TeV with the ATLAS detector}},
  {\em Phys.Rev.} {\bf D88} (2013) 112006,
  [\href{http://xxx.lanl.gov/abs/1310.3675}{{\tt arXiv:1310.3675}}].

\bibitem{Cacciari:2007fd}
M.~Cacciari and G.~P. Salam, {\it {Pileup subtraction using jet areas}},  {\em
  Phys.Lett.} {\bf B659} (2008) 119--126,
  [\href{http://xxx.lanl.gov/abs/0707.1378}{{\tt arXiv:0707.1378}}].

\bibitem{Krohn:2013lba}
D.~Krohn, M.~Low, M.~D. Schwartz, and L.-T. Wang, {\it {Jet Cleansing: Pileup
  Removal at High Luminosity}},  \href{http://xxx.lanl.gov/abs/1309.4777}{{\tt
  arXiv:1309.4777}}.

\bibitem{Berta:2014eza}
P.~Berta, M.~Spousta, D.~W. Miller, and R.~Leitner, {\it {Particle-level pileup
  subtraction for jets and jet shapes}},
  \href{http://xxx.lanl.gov/abs/1403.3108}{{\tt arXiv:1403.3108}}.

\bibitem{Randall:1998uk}
L.~Randall and R.~Sundrum, {\it {Out of this world supersymmetry breaking}},
  {\em Nucl.Phys.} {\bf B557} (1999) 79--118,
  [\href{http://xxx.lanl.gov/abs/hep-th/9810155}{{\tt hep-th/9810155}}].

\bibitem{Giudice:1998xp}
G.~F. Giudice, M.~A. Luty, H.~Murayama, and R.~Rattazzi, {\it {Gaugino mass
  without singlets}},  {\em JHEP} {\bf 9812} (1998) 027,
  [\href{http://xxx.lanl.gov/abs/hep-ph/9810442}{{\tt hep-ph/9810442}}].

\bibitem{Cheung:2005ba}
K.~Cheung and C.-W. Chiang, {\it {Splitting split supersymmetry}},  {\em
  Phys.Rev.} {\bf D71} (2005) 095003,
  [\href{http://xxx.lanl.gov/abs/hep-ph/0501265}{{\tt hep-ph/0501265}}].

\bibitem{Arvanitaki:2012ps}
A.~Arvanitaki, N.~Craig, S.~Dimopoulos, and G.~Villadoro, {\it {Mini-Split}},
  {\em JHEP} {\bf 1302} (2013) 126,
  [\href{http://xxx.lanl.gov/abs/1210.0555}{{\tt arXiv:1210.0555}}].

\bibitem{Hall:2011jd}
L.~J. Hall and Y.~Nomura, {\it {Spread Supersymmetry}},  {\em JHEP} {\bf 1201}
  (2012) 082, [\href{http://xxx.lanl.gov/abs/1111.4519}{{\tt
  arXiv:1111.4519}}].

\bibitem{Hall:2012zp}
L.~J. Hall, Y.~Nomura, and S.~Shirai, {\it {Spread Supersymmetry with Wino LSP:
  Gluino and Dark Matter Signals}},  {\em JHEP} {\bf 1301} (2013) 036,
  [\href{http://xxx.lanl.gov/abs/1210.2395}{{\tt arXiv:1210.2395}}].

\bibitem{Hisano:2006nn}
J.~Hisano, S.~Matsumoto, M.~Nagai, O.~Saito, and M.~Senami, {\it
  {Non-perturbative effect on thermal relic abundance of dark matter}},  {\em
  Phys.Lett.} {\bf B646} (2007) 34--38,
  [\href{http://xxx.lanl.gov/abs/hep-ph/0610249}{{\tt hep-ph/0610249}}].

\bibitem{Cohen:2013ama}
T.~Cohen, M.~Lisanti, A.~Pierce, and T.~R. Slatyer, {\it {Wino Dark Matter
  Under Siege}},  {\em JCAP} {\bf 1310} (2013) 061,
  [\href{http://xxx.lanl.gov/abs/1307.4082}{{\tt arXiv:1307.4082}}].

\bibitem{Fan:2013faa}
J.~Fan and M.~Reece, {\it {In Wino Veritas? Indirect Searches Shed Light on
  Neutralino Dark Matter}},  {\em JHEP} {\bf 1310} (2013) 124,
  [\href{http://xxx.lanl.gov/abs/1307.4400}{{\tt arXiv:1307.4400}}].

\bibitem{Ackermann:2011wa}
Fermi-LAT collaboration, M.~Ackermann et~al., {\it {Constraining Dark Matter
  Models from a Combined Analysis of Milky Way Satellites with the Fermi Large
  Area Telescope}},  {\em Phys.Rev.Lett.} {\bf 107} (2011) 241302,
  [\href{http://xxx.lanl.gov/abs/1108.3546}{{\tt arXiv:1108.3546}}].

\bibitem{Abramowski:2013ax}
H.E.S.S. Collaboration, A.~Abramowski et~al., {\it {Search for photon line-like
  signatures from Dark Matter annihilations with H.E.S.S}},  {\em
  Phys.Rev.Lett.} {\bf 110} (2013) 041301,
  [\href{http://xxx.lanl.gov/abs/1301.1173}{{\tt arXiv:1301.1173}}].

\bibitem{Consortium:2010bc}
CTA Consortium, M.~Actis et~al., {\it {Design concepts for the Cherenkov
  Telescope Array CTA: An advanced facility for ground-based high-energy
  gamma-ray astronomy}},  {\em Exper.Astron.} {\bf 32} (2011) 193--316,
  [\href{http://xxx.lanl.gov/abs/1008.3703}{{\tt arXiv:1008.3703}}].

\bibitem{Bergstrom:2012vd}
L.~Bergstrom, G.~Bertone, J.~Conrad, C.~Farnier, and C.~Weniger, {\it
  {Investigating Gamma-Ray Lines from Dark Matter with Future Observatories}},
  {\em JCAP} {\bf 1211} (2012) 025,
  [\href{http://xxx.lanl.gov/abs/1207.6773}{{\tt arXiv:1207.6773}}].

\bibitem{Hill:2013hoa}
R.~J. Hill and M.~P. Solon, {\it {WIMP-nucleon scattering with heavy WIMP
  effective theory}},  \href{http://xxx.lanl.gov/abs/1309.4092}{{\tt
  arXiv:1309.4092}}.

\bibitem{Cushman:2013zza}
P.~Cushman, C.~Galbiati, D.~McKinsey, H.~Robertson, T.~Tait, et~al., {\it
  {Snowmass CF1 Summary: WIMP Dark Matter Direct Detection}},
  \href{http://xxx.lanl.gov/abs/1310.8327}{{\tt arXiv:1310.8327}}.

\bibitem{Ibe:2012sx}
M.~Ibe, S.~Matsumoto, and R.~Sato, {\it {Mass Splitting between Charged and
  Neutral Winos at Two-Loop Level}},  {\em Phys.Lett.} {\bf B721} (2013)
  252--260, [\href{http://xxx.lanl.gov/abs/1212.5989}{{\tt arXiv:1212.5989}}].

\bibitem{Hall:2011aa}
L.~J. Hall, D.~Pinner, and J.~T. Ruderman, {\it {A Natural SUSY Higgs Near 126
  GeV}},  {\em JHEP} {\bf 1204} (2012) 131,
  [\href{http://xxx.lanl.gov/abs/1112.2703}{{\tt arXiv:1112.2703}}].

\bibitem{Papucci:2011wy}
M.~Papucci, J.~T. Ruderman, and A.~Weiler, {\it {Natural SUSY Endures}},  {\em
  JHEP} {\bf 1209} (2012) 035, [\href{http://xxx.lanl.gov/abs/1110.6926}{{\tt
  arXiv:1110.6926}}].

\bibitem{ArkaniHamed:2005yv}
N.~Arkani-Hamed, S.~Dimopoulos, and S.~Kachru, {\it {Predictive landscapes and
  new physics at a TeV}},  \href{http://xxx.lanl.gov/abs/hep-th/0501082}{{\tt
  hep-th/0501082}}.

\bibitem{ArkaniHamed:2006mb}
N.~Arkani-Hamed, A.~Delgado, and G.~Giudice, {\it {The Well-tempered
  neutralino}},  {\em Nucl.Phys.} {\bf B741} (2006) 108--130,
  [\href{http://xxx.lanl.gov/abs/hep-ph/0601041}{{\tt hep-ph/0601041}}].

\bibitem{Chan:1997bi}
K.~L. Chan, U.~Chattopadhyay, and P.~Nath, {\it {Naturalness, weak scale
  supersymmetry and the prospect for the observation of supersymmetry at the
  Tevatron and at the CERN LHC}},  {\em Phys.Rev.} {\bf D58} (1998) 096004,
  [\href{http://xxx.lanl.gov/abs/hep-ph/9710473}{{\tt hep-ph/9710473}}].

\bibitem{Feng:1999zg}
J.~L. Feng, K.~T. Matchev, and T.~Moroi, {\it {Focus points and naturalness in
  supersymmetry}},  {\em Phys.Rev.} {\bf D61} (2000) 075005,
  [\href{http://xxx.lanl.gov/abs/hep-ph/9909334}{{\tt hep-ph/9909334}}].

\bibitem{Feng:2000gh}
J.~L. Feng, K.~T. Matchev, and F.~Wilczek, {\it {Neutralino dark matter in
  focus point supersymmetry}},  {\em Phys.Lett.} {\bf B482} (2000) 388--399,
  [\href{http://xxx.lanl.gov/abs/hep-ph/0004043}{{\tt hep-ph/0004043}}].

\bibitem{Feng:2011aa}
J.~L. Feng, K.~T. Matchev, and D.~Sanford, {\it {Focus Point Supersymmetry
  Redux}},  {\em Phys.Rev.} {\bf D85} (2012) 075007,
  [\href{http://xxx.lanl.gov/abs/1112.3021}{{\tt arXiv:1112.3021}}].

\bibitem{Akula:2011jx}
S.~Akula, M.~Liu, P.~Nath, and G.~Peim, {\it {Naturalness, Supersymmetry and
  Implications for LHC and Dark Matter}},  {\em Phys.Lett.} {\bf B709} (2012)
  192--199, [\href{http://xxx.lanl.gov/abs/1111.4589}{{\tt arXiv:1111.4589}}].

\bibitem{Cheung:2012qy}
C.~Cheung, L.~J. Hall, D.~Pinner, and J.~T. Ruderman, {\it {Prospects and Blind
  Spots for Neutralino Dark Matter}},  {\em JHEP} {\bf 1305} (2013) 100,
  [\href{http://xxx.lanl.gov/abs/1211.4873}{{\tt arXiv:1211.4873}}].

\bibitem{Cahill-Rowley:2013dpa}
M.~Cahill-Rowley, R.~Cotta, A.~Drlica-Wagner, S.~Funk, J.~Hewett, et~al., {\it
  {Complementarity and Searches for Dark Matter in the pMSSM}},
  \href{http://xxx.lanl.gov/abs/1305.6921}{{\tt arXiv:1305.6921}}.

\bibitem{Cheung:2013dua}
C.~Cheung and D.~Sanford, {\it {Simplified Models of Mixed Dark Matter}},
  \href{http://xxx.lanl.gov/abs/1311.5896}{{\tt arXiv:1311.5896}}.

\bibitem{Griest:1990kh}
K.~Griest and D.~Seckel, {\it {Three exceptions in the calculation of relic
  abundances}},  {\em Phys.Rev.} {\bf D43} (1991) 3191--3203.

\bibitem{Profumo:2004wk}
S.~Profumo and C.~Yaguna, {\it {Gluino coannihilations and heavy bino dark
  matter}},  {\em Phys.Rev.} {\bf D69} (2004) 115009,
  [\href{http://xxx.lanl.gov/abs/hep-ph/0402208}{{\tt hep-ph/0402208}}].

\bibitem{Feldman:2009zc}
D.~Feldman, Z.~Liu, and P.~Nath, {\it {Gluino NLSP, Dark Matter via Gluino
  Coannihilation, and LHC Signatures}},  {\em Phys.Rev.} {\bf D80} (2009)
  015007, [\href{http://xxx.lanl.gov/abs/0905.1148}{{\tt arXiv:0905.1148}}].

\bibitem{Harigaya:2014dwa}
K.~Harigaya, K.~Kaneta, and S.~Matsumoto, {\it {Gaugino coannihilations}},
  \href{http://xxx.lanl.gov/abs/1403.0715}{{\tt arXiv:1403.0715}}.

\bibitem{deSimone:2014pda}
A.~De~Simone, G.~F. Giudice, and A.~Strumia, {\it {Benchmarks for Dark Matter
  Searches at the LHC}},  \href{http://xxx.lanl.gov/abs/1402.6287}{{\tt
  arXiv:1402.6287}}.

\bibitem{Boehm:1999bj}
C.~Boehm, A.~Djouadi, and M.~Drees, {\it {Light scalar top quarks and
  supersymmetric dark matter}},  {\em Phys.Rev.} {\bf D62} (2000) 035012,
  [\href{http://xxx.lanl.gov/abs/hep-ph/9911496}{{\tt hep-ph/9911496}}].

\bibitem{Cohen:2013kna}
T.~Cohen and J.~G. Wacker, {\it {Here be Dragons: The Unexplored Continents of
  the CMSSM}},  {\em JHEP} {\bf 1309} (2013) 061,
  [\href{http://xxx.lanl.gov/abs/1305.2914}{{\tt arXiv:1305.2914}}].

\bibitem{Carena:2008mj}
M.~Carena, A.~Freitas, and C.~Wagner, {\it {Light Stop Searches at the LHC in
  Events with One Hard Photon or Jet and Missing Energy}},  {\em JHEP} {\bf
  0810} (2008) 109, [\href{http://xxx.lanl.gov/abs/0808.2298}{{\tt
  arXiv:0808.2298}}].

\bibitem{Ellis:1998kh}
J.~R. Ellis, T.~Falk, and K.~A. Olive, {\it {Neutralino - Stau coannihilation
  and the cosmological upper limit on the mass of the lightest supersymmetric
  particle}},  {\em Phys.Lett.} {\bf B444} (1998) 367--372,
  [\href{http://xxx.lanl.gov/abs/hep-ph/9810360}{{\tt hep-ph/9810360}}].

\bibitem{Lindert:2011td}
J.~M. Lindert, F.~D. Steffen, and M.~K. Trenkel, {\it {Direct stau production
  at hadron colliders in cosmologically motivated scenarios}},  {\em JHEP} {\bf
  1108} (2011) 151, [\href{http://xxx.lanl.gov/abs/1106.4005}{{\tt
  arXiv:1106.4005}}].

\bibitem{Rogan:2010kb}
C.~Rogan, {\it {Kinematical variables towards new dynamics at the LHC}},
  \href{http://xxx.lanl.gov/abs/1006.2727}{{\tt arXiv:1006.2727}}.

\bibitem{Fox:2012ee}
P.~J. Fox, R.~Harnik, R.~Primulando, and C.-T. Yu, {\it {Taking a Razor to Dark
  Matter Parameter Space at the LHC}},  {\em Phys.Rev.} {\bf D86} (2012)
  015010, [\href{http://xxx.lanl.gov/abs/1203.1662}{{\tt arXiv:1203.1662}}].

\bibitem{Buckley:2013kua}
M.~R. Buckley, J.~D. Lykken, C.~Rogan, and M.~Spiropulu, {\it {Super-Razor and
  Searches for Sleptons and Charginos at the LHC}},  {\em Phys.Rev.D} (2013)
  [\href{http://xxx.lanl.gov/abs/1310.4827}{{\tt arXiv:1310.4827}}].

\bibitem{Salam:2011bj}
G.~P. Salam, {\it {Perturbative QCD for the LHC}},  {\em PoS} {\bf ICHEP2010}
  (2010) 556, [\href{http://xxx.lanl.gov/abs/1103.1318}{{\tt
  arXiv:1103.1318}}].

\end{thebibliography}\endgroup

\end{document}